# The Organization of Chaos into a Molecular Trap that Supervises Ligand-Interaction, Selection and Steric Guidance Similar to Events in Black Holes

Leroy K. Davis, Gene Evolution Project, LLC

**Abstract:** In the current study, we demonstrated that allostery transpires by entropy transfers across time-spatial scales that actualize the conception of a molecular trap that supervises ligand interaction, selection, and migration into the amphipathic groove of the 14-3-3 ζ docking protein. Ligand binding transpires by steric guidance down a multi-dimensional trap constituted of superimposed chaotic, harmonic, and electromagnetic field gradients. The individual traps exist in discrete domains governed by disparate physics interconnected by their resonance states and are subjective to damping. The highly structured molecular entanglement was genesis by the organization of white noise emitted by the anarchic motion of residues that comprised many of the features of black holes.

Keywords; molecular trap, allostery, entropy transfer, black holes

## Introduction

Allostery is a complex process where information carried by standing waves along a stretched oscillating molecular string promotes energy transference between allosteric and perturbed sites that actualize innovations in protein conformational states. The ability to precisely engineer allostery in docking proteins is germane and will appropriate the design of innovative drugs. 14-3-3 ζ isoforms focused on in the current study play pivotal roles in cellular pathways related to neurodegenerative disease and cancer. They are biologically relevant and participate in over 300 interactions. [1-4] To discover eventualities of allostery on these interactions [5], we heuristically probed gene-regulated thermodynamic parameters that oversee the materialization of an innate apparatus responsible for molecular entanglement and the electronic guidance of molecules into binding sites. Ramifications of these discoveries are significant and may propitiate the reinvention of ligand interactions by targeting small molecules to regions that regulate the formation and stability of the trap.

Our previous study revealed that a composition of helix-loop-helix domains and lack of B-sheets promotes allosteric behavior resembling energy propagation on a stretched string. Whereby coils act as allosteric hotspots, and information is routed through helices by localization of amphiphilic residues in the context of helix breakers such as serine. [5, 6] Regulation of the entropy landscape by amphiphilic residue strategic localization acquiesces genomic supervision of communication by optimizing wave transduction velocities through structural domains. Relatedly, Hasisuleyman demonstrated entropy transfer between residues [6], where some act as entropy sinks and others donors. Congruously, Keul established that forces fomented in disordered regions modulate protein function. [7] Significantly, a contemporaneous discovery disclosed that information transfers not only propagate residues but also traverse time-spatial dimensions. [5] Theoretically, protein elastic matrices are continuous collision-free media comprising infinite discrete states, whereby information transverses macroscopic to sub-electron scales. [8] Predicated on 14-3-3 ζ motion, we forecasted the subsistence of three preeminent time-spatial dimensions, macroscopic, microscopic, and sub-microscopic.

The macroscopic field captivates the conformational switch entailing structural transitions from "*open*" to "*closed*" conformers and serves as the informational driver of inter-dimensional entropic transfers. Concerted motions about interdomain coils facilitate close and long-range crosstalk between helices. They actualize allosteric free energy wells that contribute to the engenderment of an entropic entanglement that supervises ligand binding. [5] Energy wells are bounded by Z fluctuation amplitudes created by standing waves traveling the molecular matrix that captivate residue dimensionless moments; thereby, encapsulate the motion dynamic exhaustively describing convolution of the continuous elastic media respective to backbone vibration modes. Notably, concomitant studies recently published by Sala [9] and Ray [10] gave analogous accounts of structural crosstalk.

At the discontinuation of the conformational switch, domain thrusting motions observed in elastic network model motion simulations [11, 12] instantaneously stop fomenting a microscopic Hamiltonian $\psi$ characterizing shock waves in the continuous media and surrounding fluid. Translation of the standing wave Hamiltonian $\Psi \rightarrow \psi \Rightarrow \phi \pm \boldsymbol{\Omega}$ genesis structural resonances $\phi$, whereby energy transference to the surrounding fluid $\boldsymbol{\Omega}$ incites longitudinal, sound, and vortical waves. These motions actualize the microscopic disruption field creating an incoherent drive culpable for thermal fluctuations and procreation of the submicroscopic dimension. [13,14] The manifestation of resonance Hamiltonians $\phi$ by the disappearance of the macroscopic wave eventuates a rapid energy transference that transforms the motion equation to a derivation of its harmonic potential versus nascent Onsager fluctuations. [15,16] Consequentially, information propagation is subject to weaker forces such as background elastic modulus $k$ and residue interaction strength captivated by adjacency $A_{ij}$. [17] Thereby, while entropy transfer in the higher dimension is dominated by the allosteric wave Hamiltonian, at the microscopic scale, inter-dimensional communication is expedited by structural resonances supervised by the media's internal energy $d\Psi_H = dQ + dW$ in accordance to the First Law of Thermodynamics. [5,18,19]

Resonance decay $\zeta = \nabla \times \boldsymbol{\phi}$ describes the field's curl into the lower dimension, whereby evaporated energy embodies a submicroscopic Hamiltonian ҕ. [5, 20] Decay radiation traverses time-spatial scales as pressure waves emanating from oscillating stretched molecular strings whose thermal distribution forms the submicroscopic dominion. Information transfer into chaos vector space is reliant on entropy diffusion rates supervised by allosteric wave Hamiltonians. In conformity with passive diffusion processes, inter-dimensional entropy transfer depends on the concentration gradient. The disruption gradient emerges as an inverse function of the internal driving force characterized by the Onsager fluctuation theorem. [15,16] The commensurate thermodynamic parameters experience time renovation as residual waves subsisting from the macroscopic dimension gradually diminish, consequential to damping and energy exchange with the surrounding fluid.

Notably, our heuristic diffusion model disclosed that genomic organization of white noise emitted by anarchic residue motions acquiesces engenderment of an innate multidimensional molecular trap. We illustrate that the entanglement device is genesis by allosteric waves that foment disruptions in the continuous elastic media. However, its procreation transpires as a repercussion of the media's response to the standing wave. We put forth an exceptional case for allosteric engineering potential by modeling molecular trap conception and ligand interactions in a synthetic docking protein SYN-AI-1 ζ engineered in Davis [21, 22], compared to native *Bos taurus* 14-3-3 ζ. By exploring the distribution, dispersion, and diffusion of matrix disruptions circumventing backbone vibration modes, we discovered that nature forms the trapping apparatus by translating turbulence across time-spatial domains. The multidimensional trap comprises chaotic, harmonic, and electromagnetic field gradients responsible for the entrapment, selection, and electronic guidance of molecules. These superimposed intricate entanglements intersect at resonance state $X = \{\omega, \vec{r}, \emptyset\}$, implying that each system $X = \{\omega, \vec{r}, \emptyset\} \rightarrow \chi$ is an eventuality of the chaos state $\chi \cap \{\Psi, \psi, ҕ\}$ that unions with the time-spatial Hamiltonians. Consequently, the superimposed devices are simultaneously influenced by innovations in the submicroscopic field while existing in distinct domains governed by disparate physics.

## Procedures

The SYN-AI-1 ζ docking protein used in this study was simulated using a synthetic evolution artificial intelligence as described in [22]. The protein was engineered based on the "Fundamental Theory of the Evolution Force" [21]. We obtained PDB data for SYN-AI-1 ζ using the I-TASSER structure analysis tool, Yang Laboratory, University of Michigan [23], and native *Bos taurus* 14-3-3 ζ from the Protein Database. [24] To equilibrate it with the synthetic docking protein, we truncated 14-3-3 ζ from residues 215-243. We also predicted its PDB using I-TASSER. Homodimerization was analyzed utilizing COTH, a program for protein complex structure prediction by dimeric threading. [26] Whereby we analyzed protein interactions and ligand binding using Cofactor and Coach. [25] The surface convolution model (SCM) was applied to predict chaos and data projected to docking protein three-dimensional structures using USCF Chimera. [5, 27]

$$P_\zeta = \nabla(\nabla \times \phi) \Rightarrow \frac{\overbrace{\begin{vmatrix} \frac{d\Delta z}{|d\Delta z|} & \frac{d\Delta\theta}{|d\Delta\theta|} & \frac{d\Delta l}{|d\Delta l|} \\ \frac{\partial\psi}{\partial\Delta z} & \frac{\partial\psi}{\partial\Delta\theta} & \frac{\partial\psi}{\partial\Delta l} \\ \psi_{\Delta z} & \psi_{\Delta\theta} & \psi_{\Delta l} \end{vmatrix}_{A_0}}^{\zeta decay}}{K_B T} \quad (1)$$

We applied the SCM to enumerate eventualities of allosteric standing waves, microscopic free energy, and the elastic response on the inception of the molecular trap. [5] To actualize these dynamics on the organization of white noise constituting the chaos field, we applied the $\vec{r}$ vibration mode to the disruption field $S = 0.5 K_B T Log P_\zeta$ fomented by the wave captivated by pressure waves permeating the chaos field from resonating helices Eq.1. [5] Submicroscopic pressure waves $P_\zeta = \{(\nabla[\nabla \times \phi])/K_B T\}_{\partial V}$ are delineated by the thermal distribution of the energy gradient respective to the closed volume of the chaos field Hamiltonian $\zeta = \nabla \times \phi$ Eq.1. The chaos field is spatially boundless; thereby, we stipulated a closed volume $\partial V$ by projecting resonance vector space to the chaos field. Resonance $\phi = l\omega^2 \vec{r}^2$ Hamiltonians were approximated as a function of backbone vibration mode $\vec{r}$ [5] and wave angular velocity in accordance to Le's plucked molecular string Fourier solution. [28] We used five permutations to capture the entire motion dynamic of the media during convolution by standing waves Eq. (2-6).

The Fourier solutions describe backbone rotational $\Delta\theta$, tangential $\Delta z$, and stretching modes $\Delta l$ fomented by the allosteric wave. Our permutations characterize rate changes in the microscopic field Hamiltonian $\psi$ respective to the overall backbone vibration mode, consequential to the standing wave's initial velocity bout directional modes, and its final velocity at wave amplitudes. We approximated $\psi$ by modeling resonances as quantum harmonic oscillators and considering the effects of residue adjacency and alpha decay into the surrounding fluids. [5] We ignored dynamic fluid motion effects. The algorithms analyze initial wave velocities as a function of minimum energy fluctuations about helix dimensionless second-moments $M_Z$ and final wave velocities captivated by instantaneous renovations in the three vibration modes.

$$\frac{\partial\psi(\Delta l)}{\partial\Delta\theta}, where\ \psi(\Delta l) = \left\{\frac{2l^2\omega_0 sin\left(\frac{\pi an}{l}\right)}{\pi^2 n^2 a(1-a)} sin\frac{n\pi\Delta l}{l} cos\frac{n\pi\Delta l}{l}\right\} \quad (2)$$

and $\omega_0 = M_Z sin(\Delta\theta)cos(\Delta\theta)$

$$\frac{\partial\psi(\Delta z)}{\partial\Delta\theta}, where\ \psi(\Delta z) = \left\{\frac{2l^2\omega_0 sin\left(\frac{\pi an}{l}\right)}{\pi^2 n^2 a(1-a)} sin\frac{n\pi\Delta z}{l} cos\frac{n\pi\Delta z}{l}\right\} \quad (3)$$

and $\omega_0 = M_Z sin(\Delta\theta)cos(\Delta\theta)$ (3)

$$\frac{\partial\psi(\theta)}{\partial\Delta z}, where\ \psi(\theta) = \left\{\frac{2l^2\omega_0 sin\left(\frac{\pi an}{l}\right)}{\pi^2 n^2 a(1-a)} sin\frac{n\pi\Delta\theta}{l} cos\frac{n\pi\Delta\theta}{l}\right\} \quad (4)$$

and $\omega_0 = M_Z sin(\Delta z)cos(\Delta z)$

$$\frac{\partial\psi(\theta)}{\partial\Delta l}, where\ \psi(\theta) = \left\{\frac{2l^2\omega_0 sin\left(\frac{\pi an}{l}\right)}{\pi^2 n^2 a(1-a)} sin\frac{n\pi\Delta\theta}{l} cos\frac{n\pi\Delta\theta}{l}\right\} \quad (5)$$

and $\omega_0 = M_Z sin(\Delta l)cos(\Delta l)$

$$\frac{\partial\psi(z)}{\partial\Delta l}, where\ \psi(z) = \left\{\frac{2l^2\omega_0 sin\left(\frac{\pi an}{l}\right)}{\pi^2 n^2 a(1-a)} sin\frac{n\pi\Delta z}{l} cos\frac{n\pi\Delta z}{l}\right\} \quad (6)$$

and $\omega_0 = M_Z sin(\Delta l)cos(\Delta l)$

## Theory

We will build a scenario that a molecular entanglement responsible for ligand interaction, selection, and electronic guidance is genesis by a rolling turbulent entropy generated by an allosteric standing wave perturbed at the inauguration of the docking protein's conformational switch from the 'open' to 'closed' conformer. As described by Treumann, entropy requires complicated underlying dynamics that appropriate many states. That only a system consisting of many subsystems, components, and particles could occupy. [29] The innate molecular trapping apparatus is such a system that constitutes multi-dimensional vector spaces characterized by unique physics. Each molecular entanglement system comprises multiple overlaying components of backbone vibration modes encapsulating particle motions that undergo time renovation by complex thermodynamic criteria. Entropy, oscillatory, and electromagnetic field gradients forming these entanglements foment combinatorial forces that facilitate ligand binding.

While Fourier's 1878 dissertation "*The Analytical Theory of Hea*t" describes heat transfer in nature as continuous, we illustrate that entropy transfers between docking protein scalar fields transpire in multiple discrete phases. Initially, during the conformational switch, the standing wave causes a wobble and foments media disruptions encircling its turbulence that promote the heterogeneous transfer of disorder across time-spatial scales dependent on the Hamiltonian. [5,18,19] Afterwards, entropy transfer is terminated for an instant then reinstated. We based our theory on Wang [30], who modeled spiral wave turbulence in a two-layer excitable media and established that when the wave reaches the boundary, it disappears and instantaneously reinstates a homogenous state that occupies the entire medium. In the case of allostery, we propose that when an allosteric wave reaches the theoretical edge of the media defined by the termination of the conformational switch, it will attempt to reinstate a homogenous vibrational state across the protein. However, due to inhomogeneities in the continuous elastic media, its reinstatement will result in the instantaneous genesis of superimposed discrete and interacting scalar fields.

Based on the dispersion of chaotic motions about interdomain coils, and the high magnitude of chaos within a structural sink located on the dorsal side of the docking protein, there is a considerable opportunity for domain wobble at the end of the conformational switch. Thereby, the "*closed*" conformer cannot exist in a stationary stasis comprised only of overlapping residual macroscopic waves and microscopic resonances but must encompass a mode that captivates domain wobble about interdomain coils. Whereby, during reinstatement, spontaneous reallocation of its Hamiltonian will procreate four discrete scalar fields, a submacroscopic field that facilitates domain resonances $\Theta$ bout interdomain coils and promotes the organization of chaos into a molecular trap; a microscopic field comprised of coupled helix resonances $\Phi$ and dynamic fluid motions $\Omega$ in surrounding media.[5] We theorize that the remaining energy will manifest as a residual standing wave across the docking protein that reestablishes the macroscopic scalar field $\Psi$. The instantaneous nonlinear decay of the vector spaces characterized by the curl of their Hamiltonians foments a submicroscopic scalar field ↄ that is spatially boundless and whose composition is thermal energy.

As the allosteric wave propagates, its motion actualizes a rolling turbulent entropy that facilitates information transference across time-spatial scales. [5] The macroscopic disruption field genesis by the standing wave acts as an entropy donor, the submicroscopic chaos field behaves as an entropy acceptor, while the microscopic scalar field is a permissibility barrier interconnecting and synchronizing their chaotic states. Communication between dimensions transpires equivalently to coupling power proportionality according to Lyon's Law and the Clausius principle. Each vector space characterizes a disparate oscillator conjoined by the ambient Hamiltonian and attempts to obtain equilibrium. The background elastic modulus acts as a spring that couples the macroscopic and microscopic disruption fields. Whereby, the surrounding fluid coalesces superimposed disruption fields to the chaos domain **Eq. 7**. [31] Consequential to conception, the chaos field acquiesces a perpetual information sink fomenting a continual information flux traversing from higher to lower dimensions dependent on field permittivity, whereby the system can never obtain equilibrium.

To investigate the eventualities of information transference on molecular trap engenderment, we developed a heuristic diffusion model that utilizes residual plots [32,33] to detect statistical anomalies in non-linear chaos behavior disguised as standard errors. However, they characterize *"elusive information"* epitomizing entropy which acquiesces the discovery of complicated thermodynamic interactions linked to molecular trap conception that exceeds sensitivity limits of standard polynomial regressions. The data elucidate variations in entropy distributions circumventing residue backbone vibration modes. Our methodology recovers data lost consequential to inaccuracies of statistical analysis tools not tailored for application in intended scientific genres. Notably, these data captivate eventualities of thermodynamic actions and barriers imposed on time-spatial dimensions throughout entropy transference processes that would unequivocally otherwise be lost.

*Effects of Macroscopic Scale Rolling Entropy on the submicroscopic Chaos Field*

$$\langle S \rangle_{flow} \xrightarrow{\Delta} \frac{1}{2} K_B T LOG \beta \cdot \left( \overbrace{\frac{\langle S_{macro} \rangle}{N_1} \rightleftarrows \frac{\langle S_{micro} \rangle}{N_2} \rightleftarrows \frac{\langle S_{smicro} \rangle}{N_3}}^{Entropy\ Flow} \right) \quad (7)$$

$$S_{dist} = \left[\frac{\partial S_{smicro}}{\partial \vec{r}} + \frac{\partial \vec{r}}{\partial \vec{t}}\right]_\Gamma \tag{8}$$

$$Y(\vec{r}) = \frac{\partial \left\{\overbrace{\left[\frac{\partial S_{smicro}}{\partial \vec{r}} + \frac{\partial \vec{r}}{\partial \vec{t}}\right]_\Gamma}^{Experimental} - \overbrace{\left[\frac{\partial S_{smicro}}{\partial \vec{r}} + \frac{\partial \vec{r}}{\partial \vec{t}}\right]_\Gamma}^{Predicted}\right\}}{\partial \vec{r}} + \frac{\partial \vec{r}}{\partial t} \tag{9}$$

$$\chi(S_{roll}, \xi_S) = \overbrace{\left\{\frac{\partial Y(\vec{r})}{\partial S_{roll}} + \left[\overbrace{\left(\frac{\partial S_{roll}}{\partial \vec{r}} + \frac{\partial \vec{r}}{\partial \vec{t}}\right) + \frac{\partial S_{roll}}{\partial T}}^{Evolution\ rate\ \xi_S}\right]\right\}}^{Experimental} - \overbrace{\left\{\frac{\partial Y(\vec{r})}{\partial S_{roll}} + \left[\overbrace{\left(\frac{\partial S_{roll}}{\partial \vec{r}} + \frac{\partial \vec{r}}{\partial \vec{t}}\right) + \frac{\partial S_{roll}}{\partial T}}^{Evolution\ rate\ \xi_S}\right]\right\}}^{Predicted} \tag{10}$$

The study of astronomical phenomena such as Alfvén intermittent turbulence driven by temporal chaos established that the foment of chaos is dependent on wave amplitude and the dispersive parameter. [34] Using a derivative of the Navier-Stokes equation [35], Mori [36] also showed that chaos and turbulence are functions of these factors. Based on the surface convolution model [5], in molecular systems such as docking proteins, these parameters are captivated by vibration mode $\vec{r}$ that captures the totality of motion during convolution of the protein. $\vec{r}$ reflects rotational freedom around the peptide bond captured by mode $\Delta\theta$ and encapsulates backbone conformational entropy described in [37]. It also reflects pliability of the media to tangential motion characterized by mode $\Delta z$ and plasticity of the continuous elastic media to stretching as captivated by mode $\Delta l$. These motions engender the chaos field, whereby each mode foments a Boltzmann entropy as a logarithmic function of their Hamiltonian.

We described the average flow of entropy across time spatial scales in Eq. 7, where the proportionality factor $\beta$ relies on oscillator mechanical parameters delineating superimposed and interacting time-spatial dimensions. [31] According to Liu [37], entropy innovations in an open system comprise information fluxes that actualize via positive or negative disruption transferences with surroundings. Whereby time-spatial evolution of chaos consists of positive definite entropy production owing to irreversible processes within the system. In allostery, we portrayed entropy flow $\partial S/\partial t = \langle S \rangle_{flow} + \sum_n S$ across dimensions and disruption genesis over $n$ overlapping and interacting anarchic vector spaces. Boltzmann entropy of the superimposed systems was enumerated, whereby each amino acid resonance was considered a discrete thermodynamic state. [38, 39] The dispersive chaos variable was a function of submicroscopic chaos distributions circumventing the allosteric wave Fig.1a,d, represented by the tangential relationship between the chaos field and vibration mode $S_{dist} = \partial S_{sMicro}/\partial \vec{r}$ Eq.8.

The rolling entropy $S_{roll} = 0.5 K_B T Log(\pi^4 n^4 \vec{r}^2/l^3)$ captivates global perturbation and stretching of the continuous media during wave traversal as epitomized by elongational elastic modulus $k_l = \pi^4 n^4 \vec{r}^2/l^3$ [5]. We modeled eventualities of this tumultuous entropy on $\langle S \rangle_{flow} \to \chi(S_{roll}, \xi_S)$ multidimensional diffusion Eq. 10 and Fig.1c,f, where it functions as the dispersion factor. $\chi(S_{roll}, \xi_S)$ encapsulates instantaneous renovations in entropy dispersion $Y(\vec{r})$ scattering states Eq.9, Fig.1b,d, that capture tidal-like responses of the docking protein to the allosteric wave. This method is advantageous in enumerating microscopic responses and eliminates the need for analytic continuation equations. [40-45] Entropy transferences between scalar fields were extrapolated by elusive information that captivates innovations as the chaos field encircling the wave undergoes expansion and contraction. These tumultuous states renovate according to a three-dimensional evolution $\xi_S$ rate typifying modifications in the vibration mode, time remodeling of the wave, and localized temperature variances around vibration modes as the wave travels. Consequential to its reliance on media convolution prescribed by the rolling entropy, $\langle S \rangle_{flow}$ delineates the continual information exchange between the macroscopic and microscopic dimensions that encapsulates subsequent transmutation of the chaos domain into a molecular trap.

We reinvented the natural entropic entanglement by modifying allosteric waves using synthetic evolution. Consequentially, media disturbances surrounding wave turbulence increased congruently to macroscopic field disruptions. And contrary to the native state, elastic media perturbations were evenly distributed around the wave Fig.1b,e. In the native docking protein Fig.1a, the microscopic response actualized a diffusion barrier characterized by a strict minimum entropy limit $S = -2.0E+07$ that modified interdimensional vector space interactions. Constraints were overwritten in SYN-AI-1 ζ, whose standing wave overwhelmed the barrier allowing for multidirectional entropy allocations **Fig. 1d**.

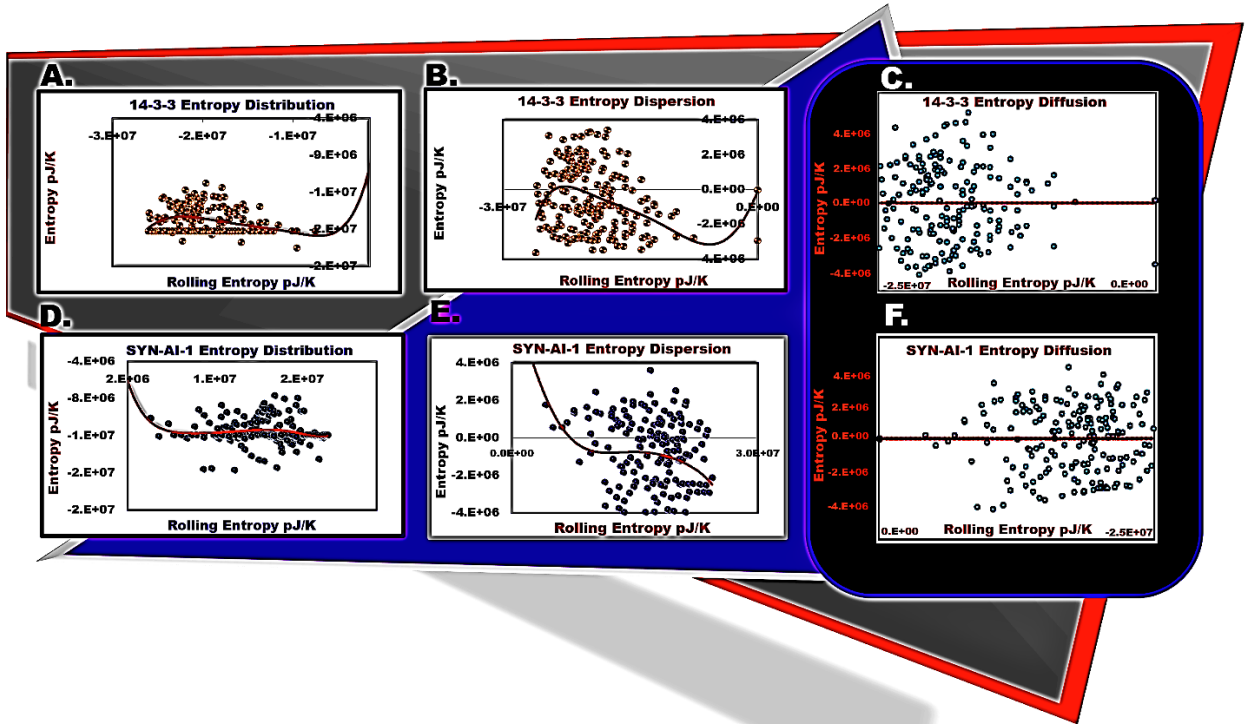

**Fig. 1 Effect of Macroscopic Rolling Entropy on the Submicroscopic Chaos Field**

*Effect of Wave Propagation on Entropic Trap*

By comparing changes in the submicroscopic scalar field in response to the vibration mode engendered by the standing wave, we discovered that the conception of the molecular trap depends on entropy transfer from the macroscopic disruption field [5] to the lower time-spatial scale **Eq. 11** and **Fig.2**. Notably, our model enabled us to demonstrate the evolution of the entropy trap by construing entropy distribution transformations around the allosteric wave.

$$\chi(\vec{r}) = \frac{\partial\left\{\overbrace{\left(\frac{\partial Y}{\partial \vec{r}} + \frac{\partial \vec{r}}{\partial \vec{t}}\right)}^{Experimental\ rate} - \overbrace{\left(\frac{\partial Y}{\partial \vec{r}} + \frac{\partial \vec{r}}{\partial \vec{t}}\right)}^{Predicted\ rate}\right\}}{\partial \vec{r}}\Bigg|_\Gamma + \left[\frac{\partial \vec{r}}{\partial \vec{t}}\right]_\Gamma \quad (11)$$

In 2016, Caro presented a paper on the underlying thermodynamics of entropy in molecular recognition by proteins [46]. We asseverate 14-3-3 ζ docking proteins manipulate the Second Law of thermodynamics [38, 47-49] by organizing chaos into a transient well-organized molecular trap actualized by actions of the vibration mode on dispersed entropy contrails **Fig.2c,f**. Notably, we innovated the nascent entanglement apparatus. The structure initially comprised an aperture that behaves similarly to an induction well fabricated of white noise emitted by chaotic residue motions (orange), a narrow funnel (yellow) interior to a palisade wall configured of white noise, and an entropic basement comprised of low chaos residue motions (red) intimately intertwined with the standing wave **Fig.2f**. The synthetic molecular trap **Fig.2c** constituted less white noise owing to information minimization repercussive to the Hamiltonian's strength. The artificial entanglement comprised a major entropy well responsible for setting the global binding potential and two minor disruption wells compared to the native molecular trap whose bulk constituted a prominent and inconspicuous well localized within aperture noise.

By actualizing the wave to interdimensional entropy transferences **Fig.2f.**, we observed the formation of a saddle that acts as the diffusion boundary mentioned in the previous section. The diffusion barrier delineates limits of residues affiliated with palisade wall formation and ligand binding from those allocated to structural stability. Localization of chaos above the diffusion saddle distinguishes white noise, whereby residues below the demarcation perform stabilizing roles. Based on their locale within the entropic bulk, residues within the lower extremity of the disruption well perform multiple tasks, including forming barrier walls, establishing the entropy

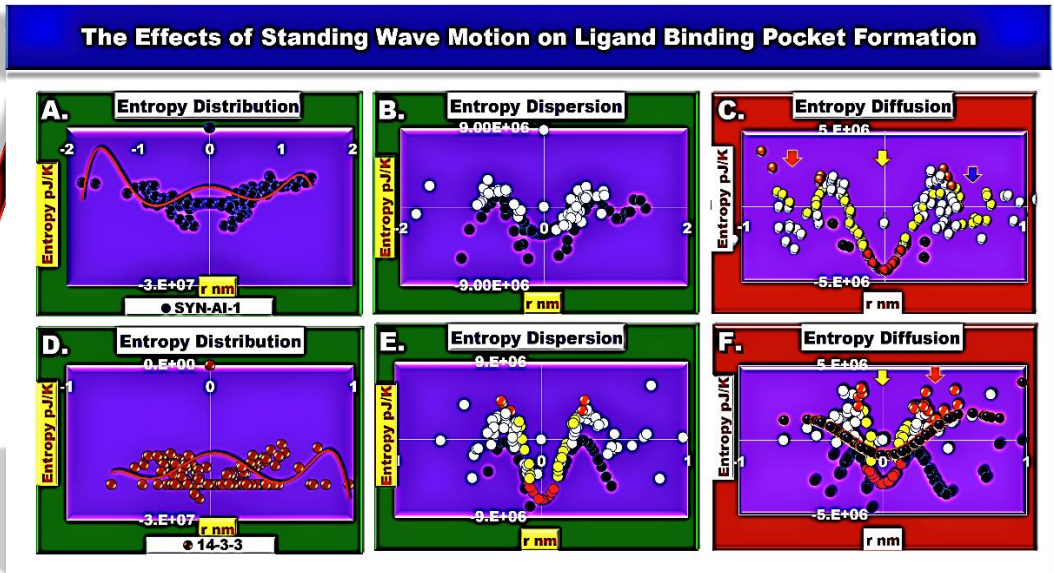

**Fig.2 Entropic Trap Formation as a Function of Standing Wave Propagation**

basement of binding sites, and structural integrity.

To corroborate Hamiltonian and vibration mode $\vec{r}$ eventualities on molecular trap formation, we subjected the SYN-AI-1 ζ chaos field to the velocity vector and actualized effects of the Hamiltonian by applying a 6th order polynomial regression to elucidate the relationship between the allosteric wave and the entanglement Fig.3. The combined influences of residue angular velocity and the macroscopic Hamiltonian fomented the molecular trapping device in regions where the Hamiltonian was weak. These locals correlated with low wave velocities and promoted entropy transference across time-spatial scales as entropy was loosely bound to the wave. Notably, the vibration mode configured chaos into an entropic snare by restricting diffusion as the vibrational radius approached the limit $\vec{r} = 0$.

We demonstrated that highly conserved residues (green) within the amphipathic groove emit white noise that facilitates entropic entanglement Fig. 4a. [50] They aid in the materialization of palisade walls and barriers implicating them in ligand binding, most informatively were not affiliated with any other region. Their role in molecular recognition was transparent, as they localized in areas that influence the groove's entropic potential, such as juxtaposed helix-coil transitions; directly within and behind binding sites Fig.4b,c. Their positions in the bulk also corresponded to a spatial depression composed solely of residues that emit white noise Fig.4a,c. This structure aids in establishing the global entropy potential by allowing flexibility during conformational switching, while interwoven low entropy residues (red) maintain structural integrity. Notably, the finding that white noise was not wasted but played a critical functional role in molecular trap formation was significant as it suggests that even at minute time-spatial scales, there is a tight genomic regulation of function.

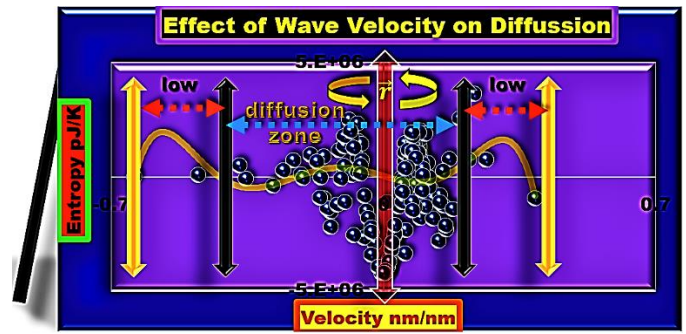

**Fig.3 Effects of Wave Velocity on Entropic Trap Formation**

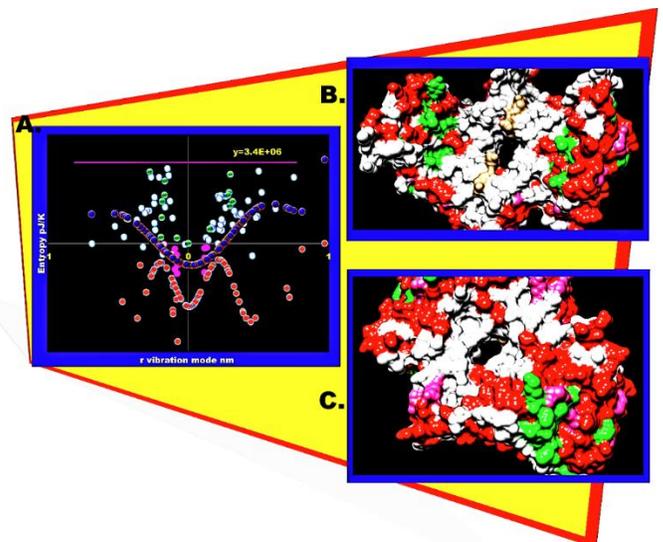

**Figure 4 White Noise Effects on Barrier Formation.**

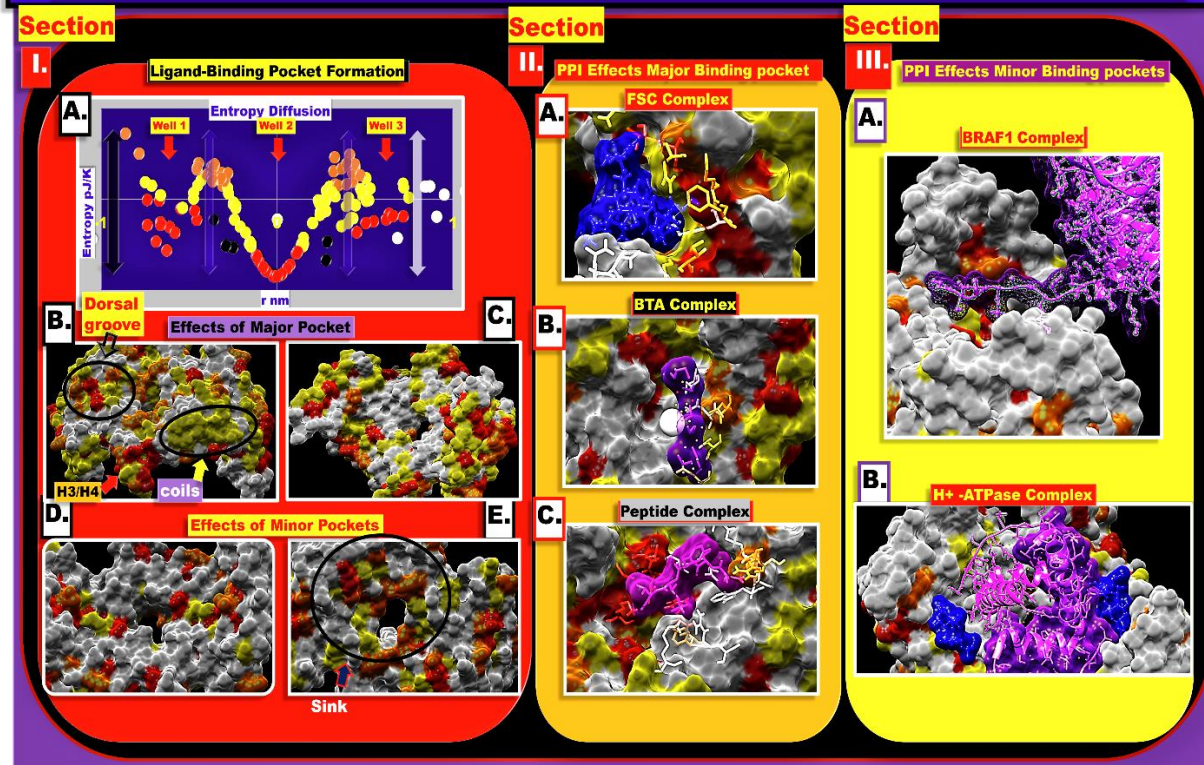

Fig. 5 The Effects of the Standing Wave on Ligand-Binding Pocket Formation

As illustrated in **Fig.5IIa**, residues constituting the major trap (well 2) of the engineered protein play a role in interactions with the fusicoccin complex (FSC, blue) that deregulates H+ ATPase activity. [51] Intriguingly, the synthetic docking protein demonstrated its potential for novel ligand interactions by binding the nitrogen-rich *N*, *N*-Bis(1*H*-tetrazole-5-yl)-Amine (BTA, purple) synthetic material **Fig.5IIb** substantiating that innovation of the entropic entanglement enhanced its function. [52] The preeminent entropic depression participated in protein-protein interactions as demonstrated by the binding of a synthetic peptide illustrated in **Fig.5IIc** (magenta). The less conspicuous traps (well1, well3) played prominent roles in ligand interactions as residues comprising these structures were intimately involved in interactions with the BRAF1 **Fig.5IIIa**, ref. [53], and the H+ - ATPase complex **Fig.5IIIb**, ref. [54].

*Effect of the Allosteric Free Energy on Entropic Trap*

While the motion dynamic about vibration mode $\vec{r}$ reflects ramifications of standing waves on entropic trap engenderment, allosteric free energy $\Delta\Delta G$ encompasses consequences at the microscopic time-spatial scale. By modeling the relationship between $\Delta\Delta G$ and the chaos field, we elucidated its eventualities on entropy gradient conception. A previous study by McLeish [20] derived $\Delta\Delta G$ in a manner that anticipated repercussions of ligand binding at an effector site on allosterically coupled locations [20]. In our previous study, we redefined $\Delta\Delta G \to \sum A e^{([\varphi+\alpha-\zeta/K_BT]+A_{ij})}$ respective to the quantum harmonic oscillation $\varphi$ to consider influences of spatial interactions epitomized by adjacency $A_{ij}$ and resonance decay $\alpha = \varphi - m\omega_0^2 \vec{r}_\Gamma^2$ transpiring within surrounding fluid as these weak interactions have a substantial influence on motion-free allostery at the microscopic scale. [5,14,17, 55,56].

Theoretically, at the end of the conformational switch, the macroscopic field motion equation $F(Z) = m(\partial^2 Z/\partial \vec{r}^2 + \Gamma\, \partial Z/\partial \vec{r}) + \nabla HMa$ decomposes to a function of its harmonic potential $\nabla HMa = F(\Psi_H) + (\phi + \Omega)$. [5,57] Whose subsequent fast decay via the sudden stopping motion transforms it to resonance $\phi \Rightarrow \varphi = (n+1/2)\hbar\omega$ and dynamic fluid motions $\Omega$. Whereby, information transfer is epitomized by allosteric wave $\hbar\omega$ traversal along a stretched oscillating string. [5]

Our derivations allowed us to corroborate the organization of chaotic motions into a molecular trap and explicate ligand-binding mechanisms. Significantly, we uncovered evidence that molecules initially interact with the mouth of a disruption well that facilitates ligand trapping and acts as a size limiter Fig. 6a. Once in the aperture, they are pulled down an entropy gradient of $6e^{07}$ $pJ/K$ in the native docking protein and $8e^{07}$ $pJ/K$ in the synthetic fomented by anarchic aperture residues and stable movements of basement residues. Notably, the subsistence of chaos, harmonic, and electromagnetic entanglements as superimpositions of resonance state $X = \{\omega, \vec{r}, \emptyset\}$ provided a theoretical basis to enumerate disruption forces according to Krishtalik [58] and Wolski [59]. Who modeled electromagnetic field intensity fluctuation precipitated by ligand-protein interactions utilizing Maxwell's Dynamic Theory of Electromagnetic Fields [60, 61] and Stoke's Theorem [62].

$$\S = \left\{ \int_{\partial V} dA \int_{\rho(0)}^{\rho(1)} \nabla(\Delta S) \cdot d\rho(\Delta\Delta G) \right. \\ \left. + \int_{\partial A} dl \int_{\sigma(0)}^{\sigma(1)} \nabla(\Delta S) \cdot d\sigma(\Delta\Delta G) \right\}_{X(\Gamma)} \quad (12)$$

$$Y(\S) = \sum_{r=0}^{\infty} \sum_{d=1}^{4} \sum_{n=0}^{\infty} \frac{1}{8\varepsilon \pi^{d+1} r^2} \left\{ \int_{\partial V} dA \int \nabla[(\mathfrak{X}_S) \cdot \mathcal{F}\{\rho(\Delta\Delta G)\}] \right. \\ \left. + \int_{\partial A} dl \int \nabla[(\mathfrak{X}_S) \cdot \mathcal{F}\{\sigma(\Delta\Delta G)\}] \right\}_n \quad (13)$$

$$\mathcal{F}\{\rho, \sigma\} = \frac{1}{2\pi^d} \left\{ \int_{\partial V} dS \int d^d k_d e^{ik \cdot \Delta\Delta G_0^d - Dk_d^2 t} \right. \\ \left. + \int_{\partial S} dl \int d^d k_d e^{ik \cdot \Delta\Delta G_0^d - Dk_d^2 t} \right\} \quad (14)$$

Entropic force § was solved over the closed volume of the docking protein and force contributed by local disruption gradients within binding sites by their line integral. We substituted entropic potential $(\Delta S = S_{aper} - S_{base})$ for electric field intensity $\varphi = \nabla \vec{E}$ as the states are exact superimpositions of the motion state Eq.12. Due to a tidal-like response to the allosteric wave, the organization of white noise to the molecular trap is dependent on $\Delta\Delta G$. Entropy force § [29] is epitomized by the divergence of the chaos gradient respective to $d\rho(\Delta\Delta G)$ the probability density of residues within the bound volume $\partial V$, and to $d\sigma(\Delta\Delta G)$ density of residues in the bound area $\partial A$ of the docking protein. Whereby, the term $(\Delta S) \cdot d\rho(\Delta\Delta G) \Rightarrow 1/p(\Psi)$ is an inverse expression of wave energy density. Expansion of the force field $\S(r, \varepsilon) = \S/4\pi\varepsilon r^2$ into the amphipathic groove depends on the distension radius and the permittivity $\varepsilon$ of free space. Entropic entanglement $Y(\S)$ is an actualization of the fluctuation of probability density states $p(\vec{r}, \Delta\Delta G) \to \mathcal{F}\{\rho(\Delta\Delta G)_n\}$ over time and space $r$. Dimensions $d \to +1$ describes four-dimensional spacetime and $d$ depicts the time-spatial scales Eq.13. $\mathfrak{X}_S$ captivates the interaction of ligand and protein entropy fields and is the resultant potential. The Fourier is given by Eq.14, where $D$ is the diffusion coefficient and wave number $k = 2\pi n/\lambda$.

$$S_{Prot} = -\frac{1}{2} K_B T \int_{d=1}^{4} Ln \left\{ \frac{kA_{ij}}{(\vec{r}^2)^d} + \frac{G_n M}{|\vec{r}|^d} \right. \\ \left. + \sum_{l=2}^{\infty} \frac{1}{l} \frac{(2l+d-2)!!}{d!!} \frac{Q_L n^L}{|\vec{r}|^{l+d}} \right\} \quad (15)$$

When considering entropy potentials $\mathfrak{X}_S$ experienced by the interacting ligand and docking protein, we solved according to Keinbaum [40] and applied a Boltzmann entropy transformation to convert allosteric free energy $\Delta\Delta G$ to anarchic motion. Additionally, we considered contributions by the continually evolving internal energy vector space Eq. 15. The first term captivates effects of modulating residue interaction strengths $A_{ij}$ transpiring during media convolution and eventualities of a dynamically innovating elastic moduli $k$ on microscopic field Hamiltonian $\Delta\Delta G = kA_{ij}/(\vec{r}^2)^d$. The second term accounts for gravity impacts, and the third for magnetic influences. Whereby, $\mathfrak{X}_S = \Delta S_{prot} + \Delta S_{ligand}$ captures the combinatorial effects of the interacting entropy gradients consequential to the localization of the ligand within the entanglement and its orientation. $G_n$ is the d-dimensional gravitational constant, $l$ multipolar order, and $M$ molecular mass.

D-dimensional Newtonian moments $Q_L$ result from vibration mode $\vec{r}'$ innovation, where Fourier Transform $\rho(\Delta\Delta G) \Rightarrow \mathcal{F}\{\rho, \sigma\}$ describes volume-surface probability densities respective to time and the initial position.

$$Q^L = \int d^{d+2} \vec{r}' \, \mathcal{F}\{\rho, \sigma\} \vec{r}'^L \quad (16)$$

Interaction effects on the extended object were also modeled according to Keinbaum Eq.17. [40] However, with the consideration of internal energy influences on the tidal tensor $\varepsilon_L$ consequential to the average allosteric free energy potential $Д = \langle i \cdot \Delta\Delta \hat{G}_{prot} + j \cdot \Delta\Delta \hat{G}_{lig} \rangle$ that transpires as a function of the Hamiltonian direction depicted by unit vectors $(i, j)$; thereby accounting for the peristaltic counteracting motion gradients that propel molecules down the bulk Eq 18. Whereby, radius $r^{l+d}$ reports the instantaneous distance between the protein and ligand. Although, mass effects vary with local three-dimensional structure and ligand localization within specific binding sites, the molecule's motion within the trapping apparatus can be mapped to the dimensional distortion, and radius $r^{l+d}$ determined by its bulk Fig. 6.

$$S_{prot}^{ext} = -\frac{1}{2} K_B T Ln \sum_{l=0}^{\infty} \frac{1}{l!} \vec{r}^L \varepsilon_L \tag{17}$$

$$\varepsilon_L = \frac{M_{lig} G_n}{r^{l+d}} (-1)^{l+1} \frac{(2l+d-2)!!}{d!!} n_p^L \cdot Д \tag{18}$$

The highly structured entropic entanglement Fig.6a is genesis by the combined actions of scalar fields, where the continuous elastic media's elastic response confines disorder fomented by the macroscopic wave and microscopic structural resonances. The chaos force is actualized by an entropy potential between the higher time-spatial scales and the submicroscopic scale that acts as a dimensionless, spatially infinite entropic sink that promotes the transport of excess entropy across fields. Due to the rapid dissipation of thermal energy out of the sink, the trap should instantaneously dissipate. However, it is stabilized by a constant refurbishing of entropy from residual standing waves and structural oscillations that supply the chaos factor. The temporal nature of the trap allows it to manipulate the Laws of Thermodynamics, while the organization of chaos into structure sets the selectivity for ligand interactions. Combined actions of downward and concentric chaos gradients about the $\Delta\Delta G$ scalar field sterically guide ligands into a specific binding site through a well-ordered isentropic disruption funnel Fig.6a, whose acclivity is V-shaped Fig. 6b in the synthetic docking protein due to a more vigorous elastic response that reallocates function.

We corroborated our presumption of steric guidance by demonstrating that chaos was not only associated with motion but displayed a spatial relationship in the protein structure relative to its location on the entropic trap Fig.6c-f. Residues at the extreme basement (black) of the preeminent well were consistently located at the base of binding sites, while white noise surrounding the energetic pocket formed walls of the amphipathic groove that helped secure ligands in place. The tendency was observable in the interaction of the small molecule (2S) - (2) - methoxyethyl pyrrolidine [22] with the molecular entanglement device Fig.6g. Whereby aperture residues (orange) localized at the top of binding sites, funnel residues (yellow) and white noise formed ligand-binding site walls. Entropy floor residues (red) of the small well superimposed within the ostentatious one, placed spatial limitations on the size of molecules that can appropriate sites. Notably, the discovery of congruous relationships in interactions involving FSC complex Fig.6h and small peptides Fig.6i,j, corroborated these findings.

$$\chi(\Delta\Delta G) = \left\{ \frac{\partial \chi}{\partial \Delta\Delta G} + \overbrace{\frac{\partial \Delta\Delta G}{\partial A_{ij}} \cdot \frac{\partial A_{ij}}{\partial k} \cdot \frac{\partial k}{\partial \vec{r}} \cdot \frac{\partial \vec{r}}{\partial t}}^{Evolution\ rate\ \xi_{\Delta\Delta G}} + \overbrace{\frac{\partial \Delta\Delta G}{\partial \alpha} \cdot \frac{\partial \alpha}{\partial \vec{r}} \cdot \frac{\partial \vec{r}}{\partial t}}^{Evolution\ rate\ \xi_\alpha} \\ + \frac{\partial \Delta\Delta G}{\partial \zeta} \right\}_\Gamma \tag{19}$$

Entropic entanglement is a function of entropy scattering dependent on diffusion transference rates $\chi$ across time-spatial dimensions and consequential to the instantaneous slope of these rates respective to available $\Delta\Delta G$ within the elastic media Eq.19. $\chi(\Delta\Delta G)$ vector space comprises a complicated multifactorial evolution operator encompassing residue interaction strength $A_{ij}$ renovations and $\alpha$ decay of the resonance Hamiltonian [5]. Information transfer is likewise influenced by heat loss $\nabla \times \zeta = \nabla \times (\nabla \times \Delta\Delta G)$ characterized by the curl of the submicroscopic field Hamiltonian $\zeta$.

### Effects of the Elastic Response on Entropy Diffusion

We deduced that behavioral differences in synthetic and nascent entropic entanglements $Ұ(\S)$ owed to their elastic responses $\psi_\Gamma \to A_{ij}k = \Delta\Delta G/\vec{r}^2$ to the standing wave and actualized by combined influences of residue interaction strength and order in the continuous media Eq. 20. The intercorrelation embodies actualities of the allosteric wave Hamiltonian at the microscopic scale and reciprocated repercussions of the elastic media on the wave respective to the vibration mode it inspires. Much like waves traversing a fluid, a strong Hamiltonian associated with a low density, low viscosity liquid will foment a significant amount of turbulence and entropy dispersion or splash. Equally tumultuous waves traveling in a dense and high viscosity fluid will smoothly traverse the media displaying succinct chaos as the elastic response will set strict energetic boundaries that serve as a container. We observed this tendency in the first chapter as the ER formed a barrier that prevented disruption transfers across time-spatial scales, thereby regulating entropy allocations and the organization of white noise into an entropic trap.

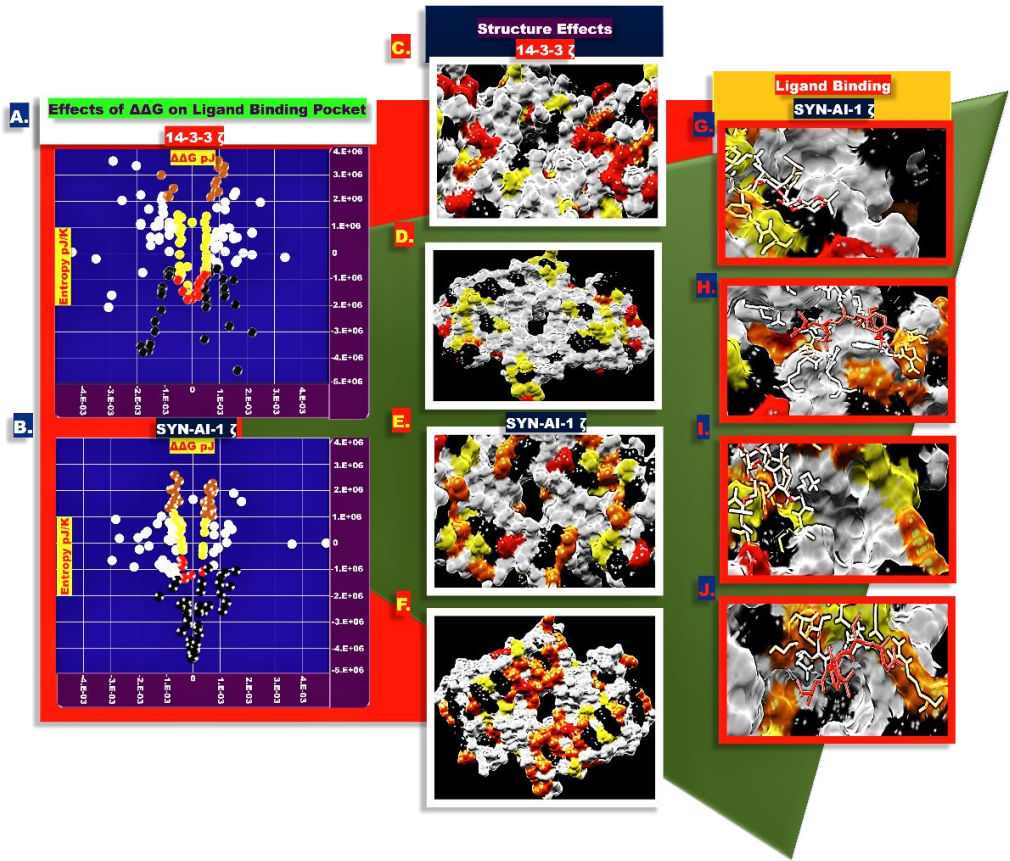

**Fig.6 The Effect of Allosteric Free Energy on Entropy Diffusion and Formation of the Entropic Trap**

$$\psi(\Gamma) = \overbrace{\widehat{kA_{ij}}}^{micro} \Rightarrow \left\{ \overbrace{\frac{\partial \Delta\Delta G}{\partial S_{smicro}^2}}^{\overbrace{Evolution\ rate\ \xi_\psi}^{smicro}} + \left(\frac{\partial S}{\partial T} + \frac{\partial T}{\partial P}\right)^2 + \left(\frac{\partial S}{\partial \zeta} + \frac{\partial \zeta}{\partial \Delta\Delta G}\right)^2 \right\}_\Gamma \quad (20)$$

$$\chi(\beta_\psi) = \frac{\partial \chi(\vec{r})}{\partial \beta_{\psi(\Gamma)}} + \xi \quad (21)$$

$$\left\{ \left( \overbrace{\frac{\partial \beta_{\psi(\Gamma)}}{\partial A_{ij}} + \frac{\partial A_{ij}}{\partial k} + \frac{\partial k}{\partial \vec{r}} + \frac{\partial \vec{r}}{\partial \vec{t}}}^{\xi_{A_{ij}}} \right) + \left( \overbrace{\frac{\partial \beta_{\psi(\Gamma)}}{\partial k} + \frac{\partial k}{\partial \vec{r}} + \frac{\partial \vec{r}}{\partial \vec{t}}}^{\xi_k} \right) + \overbrace{\frac{\partial \beta_{\psi(\Gamma)}}{\partial \zeta}}^{\xi_\zeta} \right\}_\Gamma \quad (22)$$

To demonstrate ramifications of $\psi_\Gamma$ on molecular trap engenderment, we modeled the information transference rate Eq.21 across time-spatial dimensions respective to the ER barrier $\beta(\psi) = 0.5 K_B T Log\ \psi_\Gamma$. We actualized $\beta(\psi)$ onto the instantaneous slope of the chaos vector space eventuated by the allosteric wave as captivated by $\chi(\vec{r})$, Fig.7a,b. Consequently, enacting an extra white noise barrier to the naturally transpiring ER of the 14-3-3 ζ docking protein. The evolution ξ constant for the entropy barrier $\beta_{\psi(\Gamma)}$ is a function of dynamic residue adjacency $\partial A_{ij}$ and elastic modulus $\partial k$ states that innovate subjective to decay ζ Eq.22.

By applying the ER barrier, we discovered that while entropy transfer across time-spatial dimensions was continuous, it transpired within three distinct domains supervised by $\psi_\Gamma$ that acquiesce diverse effects Fig. 7a. Inaugurally, it acted as a permissibility barrier whereby diffusion initially transacted in the domain delimited by $\psi_\Gamma = 0$. This domain correlates with helix-loop-helix motions about interdomain coils and actualizes crosstalk between structures that cooperatively genesis the entropic trap. Succeeding, the ER created a partition that acquiesced a restriction zone where chaos transference $\chi(\beta(\psi)) = \emptyset : \{0 < \psi_\Gamma < 1.84E^4\, pJ/K\}$ across time-spatial scales was null. The partition acceded genomic allocation of function and molecular trap materialization in the third domain. Incipiently, $\psi_\Gamma$ regulated the trap's length by stretching the chaos field outwardly **Fig.7a,b**.

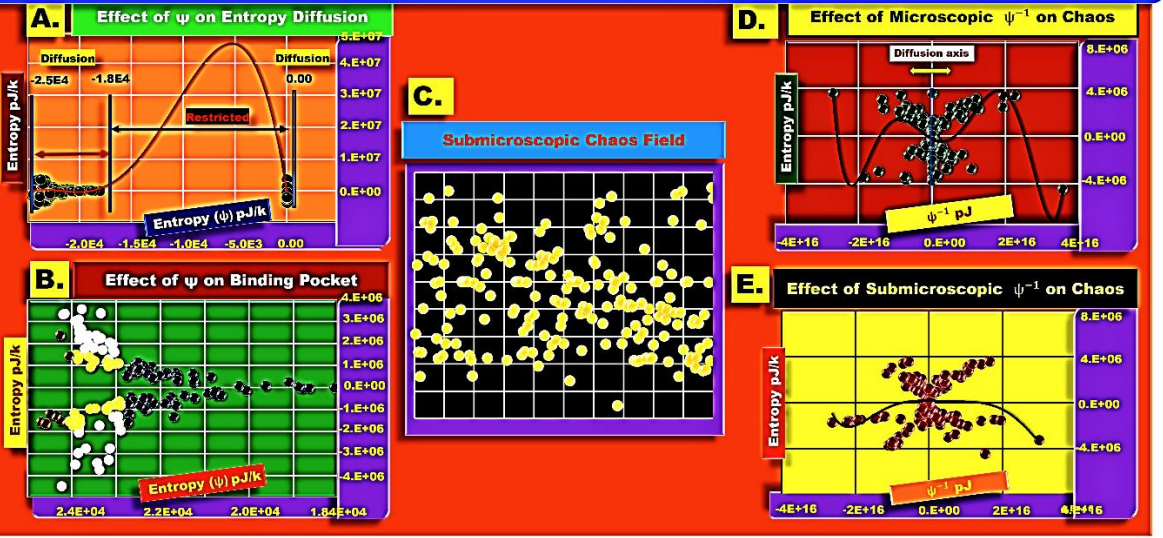

**Fig.7** The Effects of the Elastic Response on Ligand-Binding Pocket Formation

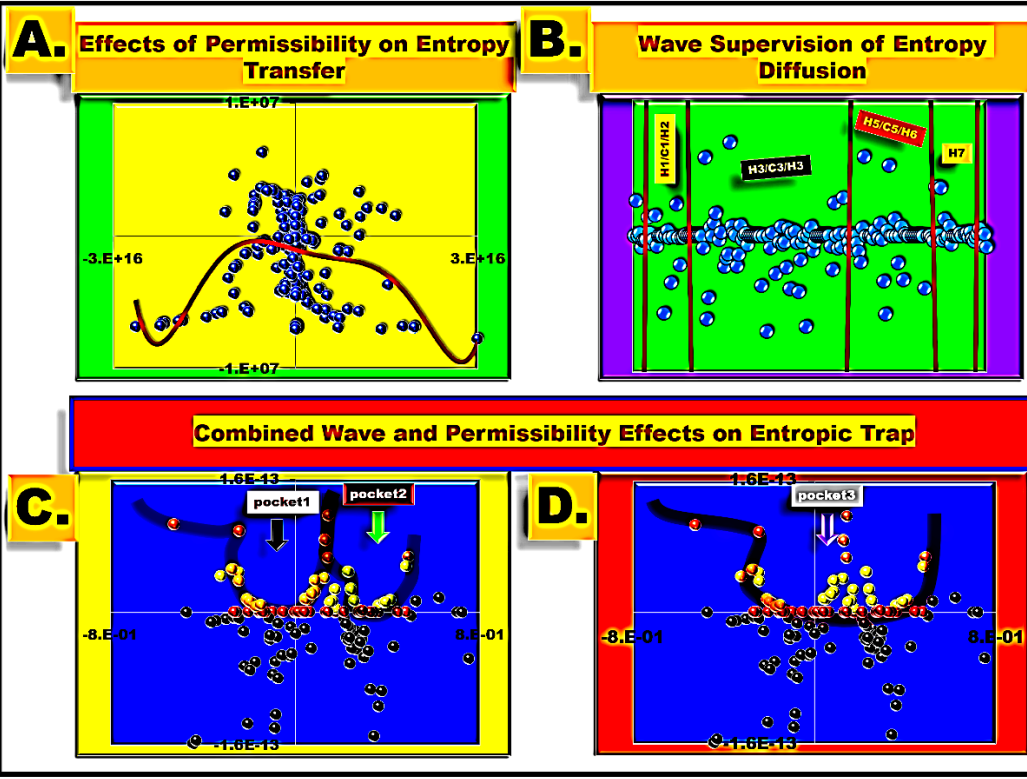

**Fig.8** The Effect of Permissibility on Entropic Trap Formation

Notably, our application of a second ER barrier resulted in the reallocation of function we espied in the synthetic docking protein.

Additionally, information transference is contingent on elastic media permissibility $\psi_\Gamma^{-1}$ which dictates entropy gradients Fig.7d at the microscopic scale and relies upon chaos field permissiveness $\psi_\Gamma^{-1}$ to incoming disruptions Fig.7e. The allosteric free energy gradient $\nabla(\Delta\Delta G) = \partial\Delta\Delta G/\partial\psi_\Gamma$ explicates media permissibility characterized by $\Delta\Delta G$ vector field instantaneous slope respective to the media's response to the allosteric wave. Its actualization to chaos vector space scrutinizes these eventualities onto the lower dimension Fig.7d. Whose permittivity $\varrho = \partial\chi/\partial\psi_\Gamma^{-1}$ is given by the transference rate respective to the dimensional permissibility Fig.7e. Because the submicroscopic chaos field is spatially boundless, we modeled the ER $\psi_\Gamma = \Delta\Delta G/S_{smicro}^2$ in the thermal dimension as an implicit expression of media internal energy $\Psi = dQ + dW$ by mapping the vibration mode $\vec{r} \to S_{micro}$ to the chaos field. The terms are thermodynamically synonymous; the latter acquiesces profound aftermaths of evaporation on the system.

$$\chi(\beta(\psi)) = \left\{\frac{\partial\left[\overbrace{\left(\frac{\partial Y}{\partial\psi_\Gamma^{-1}} + \xi_S\right)}^{\text{Experimental rate}} - \overbrace{\left(\frac{\partial Y}{\partial\psi_\Gamma^{-1}} + \xi_S\right)}^{\text{Predicted rate}}\right]^{\text{Residual Entropy}}}{\partial\beta(\psi)}\right\}_\Gamma \quad (23)$$

To heuristically probe the intricacies of entropic entanglement, we analyzed disruption transference into the chaos field based on the instantaneous dispersion slope respective to microscopic scale permissibility $\psi^{-1}$ Fig.8a. We considered combinatorial eventualities of permissibility and the ER barrier on the transference of anarchical behavior by applying a diffusion barrier $\beta(\psi) = 0.5K_BTLog(\Delta\Delta G/\psi)$ to the dispersed residual entropy Eq.23. Eventualities of the allosteric wave on multi-dimensional entropy exchanges were ascertained by the wave's relationship to submicroscopic residual entropy circumventing its turbulence by applying a 6th order polynomial regression Fig. 8b. Entropy transfer transpired at the domain level, whereby interdomain coils formed wave nodes and surrounding helix-loop-helix structures antinodes. By actualizing the $\vec{r}$ vibration mode onto the diffusion vector $\chi(\beta(\psi))$, we isolated entropy depressions reported in initial experiments Fig.8c. Notably, these entropic apparatuses interact as implied by their juxtaposed positioning within the more ostentatious well that encapsulates the global trapping capacity of the entanglement Fig.8d.

We discovered that the entropic entanglements were intimately associated with ligand binding sites Fig.9a,b. Entropy floor residues (red) were distinguishable from funnel residues (yellow) and affiliated with interdomain coils and helix-coil transitions that display coil-like behavior, corroborating their role in global entropy potential supervision. We verified that funnel-residues regulate local disruption gradients as they localized to helix H3 and H4 hinge-bending areas Fig.9e (arrows) that juxtapose binding sites. We validated their influence on the entropic potential by superimposing their motion to the localization of amphiphilic residues (magenta) that induce local disorder Fig.9b,c,e,f. [5] Noticeably, they were exclusively amphiphilic or localized contextually in a sequence of amphiphilic residues that persistently also included alanine, whose small volume allows rotational freedom or serine, a well-established helix breaker. [5] Thereby, we corroborated that funnel-residues supervise entropic potentials within the amphipathic groove. The finding was also supported by their localization directly behind binding sites depicted by circled regions.

*Effects of the Energy Gradient on Entropy Diffusion*

Our findings imply that strategic localization of residues permits the elastic response to modulate function by supervising the organization of chaos, enabling ligand interaction and migration to be genomically regulated and evolutionarily optimized. We substantiated macroscopic force field eventualities by actualizing them unto the chaos field Fig. 10a, whereby the Onsager fluctuation theorem predicted force [5]. We corroborated ER barrier supervision of entropy transference by subjecting the macroscopic disruption field characterized by Boltzmann entropy $S = 0.5K_BTLog(F)$ to the ER barrier and reconstructing the molecular trap Fig. 10d. Allosteric wave diffusion $\chi(\psi, \vec{r}) = \partial S^2(F)/\partial \psi \partial \vec{r}$ was expounded respective to the elastic response $\psi$ **Fig.10b**. Combinatorial repercussions of higher-order time-spatial scales on turbulence translations into the submicroscopic field were captivated by the anarchic thermodynamic relationship $\chi(\psi, \vec{r}) = \partial \chi_{smicro}/\partial \chi(\psi, \vec{r})$.

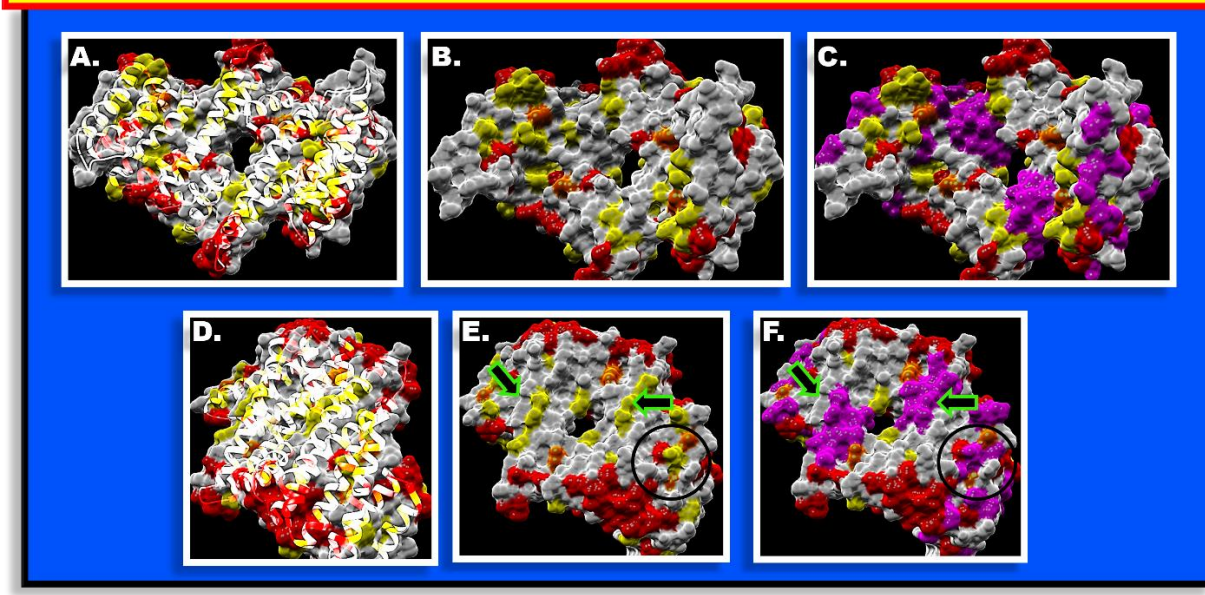

**Fig.9 The Effect of Permissibility on Ligand-Binding Pocket Formation.**

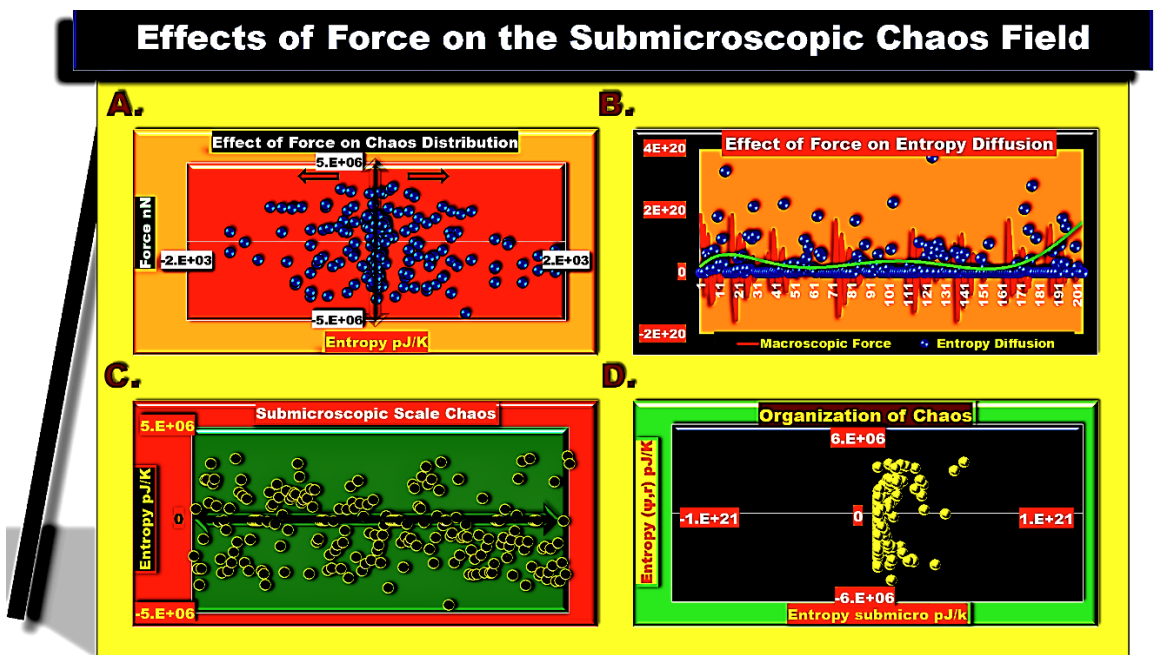

**Figure 10. The Effect of Force and the Elastic response on Organization of Chaos into Structure**

The ER acts on available entropy determined by the diffusion gradient **Fig. 10a**. Entropy transferences across time-spatial scales were distributed around a driving force of $F = 0$, implying the gradient actuates inversely to the force field. Low-intensity force fields were associated with helix-loop-helix domain motions about allosteric wave nodes and regions where allostery was routed through helices [5], **Fig.10b**. We corroborated the strict disruption barrier recognized beforehand was eventuated by the ER, as we reproduced the effect in **Fig.10d**. We expatiated that the ER barrier institutes the overall diffusion limit. However, the allosteric wave establishes diffusion limits for white noise. Our findings demonstrate that combined effects of the macroscopic force field and the continuous elastic media's response at the microscopic scale are sufficient to distort the thermal dimension and organize the chaos field **Fig.10c** into a molecular trap **Fig.10d**.

*Microscopic field Damping and Trap Stability*

Our preliminary analysis suggests the transient interaction and release of ligands by an innate molecular entanglement apparatus is driven by chaos organization propitiated via a microscopic elastic response that interfaces macroscopic and submicroscopic scales. We construed the repercussions of damping on entropic entanglement stability responsive to inter-dimensional transference rates $\chi(\Delta\Delta G) = \Lambda_i(\partial \chi/\partial \Delta\Delta G)$ by applying evolution operator $\Lambda_i$ to the instantaneous field slope in relationship to allosteric free energy $\Delta\Delta G$. $\Delta\Delta G$ dictates the conveyance of turbulence into lower time-spatial scales, and its decay supplies spontaneity for chaos field engenderment. Application of evolution operator $\Lambda_0 = \sum X' \frac{\partial f(x)}{\partial X}$ implies innovation of nascent state $f(x)$ correspondent to a parameter $X$ is consequential to derivative $X'$; which is a member of a domain **Eq.25** of multidimensional operators $\xi_S$ **Eq.24** formed by linear chains of congruous thermodynamic criteria.

$$\Delta\Delta G \xrightarrow{\Lambda_1} \overbrace{\sum A'_{ij}\frac{\partial \Delta\Delta G}{\partial A_{ij}}, A_{ij} \xrightarrow{\Lambda_2} \sum k'\frac{\partial A_{ij}}{\partial k}, k \xrightarrow{\Lambda_3} \sum_{j=1}^{k} \vec{r}'\frac{\partial k}{\partial \vec{r}}, \vec{r} \xrightarrow{\Lambda_4} \sum_{j=1}^{k} t'\frac{\partial \vec{r}}{\partial t}}^{Linear\ events} \quad (24)$$

$$\chi(\Delta\Delta G) \xrightarrow{\Lambda} \sum \frac{\partial \chi}{\partial \Delta\Delta G} \cdot \overbrace{\left\{\begin{array}{c} \overbrace{\frac{\partial \Delta\Delta G}{\partial A_{ij}}\frac{\partial A_{ij}}{\partial k_j}\frac{\partial k}{\partial \vec{r}}\frac{\partial \vec{r}}{\partial t}}^{\Lambda_S(A_{ij})} + \overbrace{\frac{\partial \Delta\Delta G}{\partial \alpha}\frac{\partial \alpha}{\partial \vec{r}}\frac{\partial \vec{r}}{\partial t}}^{\Lambda_S(\alpha)} \\ + \underbrace{\frac{\partial \Delta\Delta G}{\partial A_{ij}(ligand)}}_{\Lambda_S(Ligand)} + \underbrace{\frac{\partial \Delta\Delta G}{\partial \alpha(Ambient)}}_{\Lambda_S(Ambient)} \end{array}\right\}}^{\Lambda_{\Delta\Delta G}} \partial t \quad (25)$$

The allosteric free energy landscape evolution operator $\Lambda_{\Delta\Delta G}$ is dependent on residue interaction strength $A_{ij}$ and energy exchange $\alpha$ with surrounding fluid [5]. Innovations are predisposed to ligand-bound and unbound protein configurations that alter local and global resonance states and are also influenced by the ambient temperature. The molecular entanglement state is an exponential growth nonlinear differential equation $\chi(\Delta\Delta G)_{t+1} = \chi(\Delta\Delta G)_t \exp[\Lambda(1 - \chi(\Delta\Delta G)_t)]$ [63] that encaptivates diffuse distributions of white noise. The implicit form characterizes its probability $p(X,t) = \left(\pi\sqrt{\chi(\Delta\Delta G)_t(1 - \chi(\Delta\Delta G)_t)}\right)^{-1}$ density where the evolution rate is $\Lambda_{prob} = \partial p(X, t+1)/\partial p(X, t)$. [64] The continuity equation $\{\partial p(X,t)/\partial \vec{r}\} + \nabla \cdot J_{flux} = 0$ implies that distribution innovations $\partial p(X,t)/\partial \vec{r} = D\nabla^2 p(\vec{r}, t)$ are consequential to the diffusion constant. In molecular systems, information transfer $\nabla p(\vec{r}, t) = 1/\nabla F$ is attributed to local residue density and inverse gradients of force fields. Multidimensional information $J_{flux} = \psi^{-1}\Lambda_{prob} - \Gamma\chi_{\vec{r}}e^{(1-\chi(\vec{r})_t)}\frac{\partial}{\partial \vec{r}}\Lambda_{prob}$ flux across time-spatial scales depend on the permissibility of interacting disruption fields, their relative densities, and innovation of the diffusion rate $\chi_{\vec{r}}$ correspondent to the vibration mode and damping $\Gamma_{rate} = \Delta\Delta G_{t+1}/\Delta\Delta G_t$. Disruption radiation $\varrho(t) \Rightarrow p(X, t)$ into space is given by the Fourier Transform **Eq. 26** over radius $r$, time-spatial scales $d$, and diffusion states $n$. In the spatial dimension, we used Laplace Transform $\varrho(\vec{r}) \Rightarrow \mathcal{L}\{p(\vec{r}, t)\} = \tilde{p}(\vec{r}, s)$. Where the nascent state in $d$ dimensions is depicted by $p(\vec{r}, X_0) = \delta^d(\vec{r})$, the entropy diffusion given according to $s\tilde{p}(\vec{r}, s) - p(\vec{r}, X_0) = D\nabla^2 p(\vec{r}, s)$, and initial boundary conditions by $(s - D\nabla^2)\tilde{p}(\vec{r}, s) = \delta^d$. [65]

$$\varrho(t) = \sum_{r=0}^{\infty}\sum_{d=1}^{4}\sum_{X=0}^{n}\frac{1}{8\varepsilon\pi^{d+1}r^2}\left\{\int_{\partial V} dS \int d^d k_d e^{ik\cdot X_0^d - Dk_d^2 t} + \int_{\partial S} dl \int d^d k_d e^{ik\cdot X_0^d - Dk_d^2 t}\right\}_n \quad (26)$$

$$\chi(\Delta\Delta G) = \frac{\partial \chi}{\partial \Delta\Delta G}\sum_{\xi_S=1}^{n}\prod_{a=1}^{m} y_a = \int_{y>saddle}^{y=upper}\chi(\Delta\Delta G)dt \quad (27)$$

The decay Hamiltonian also experiences multiple states induced by transient temperature flux about backbone vibration modes during wave propagation. These states are probability $\Lambda_S = \prod y_a$ functions of thermodynamic criterion $y = \partial/\partial x$ that govern motion and energy exchange **Eq. 27**. Whereby the summation of entropies localized above the diffusion saddle is sufficient to describe the chaotic state. Meanwhile, lower limit $y = \{ax + bx^2 + cx^3 + dx^4 + dx^5+ \dots\dots\}$ is given by a multivariable polynomial regression derivatized as a function of the nascent diffusion state $\partial \chi/\partial \Delta\Delta G$. Limit $y$ simulates allosteric wave interactions with the docking protein during the conformational switch **Fig. 11**.

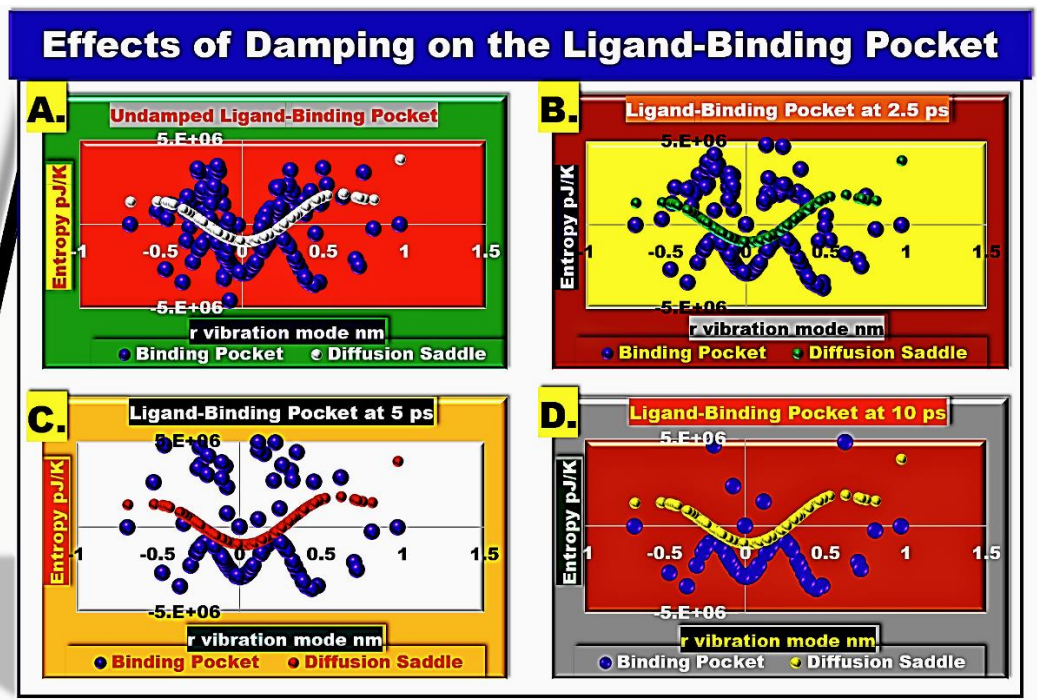

**Fig. 11.** Deconfiguration of the Molecular Trap Due to Damping

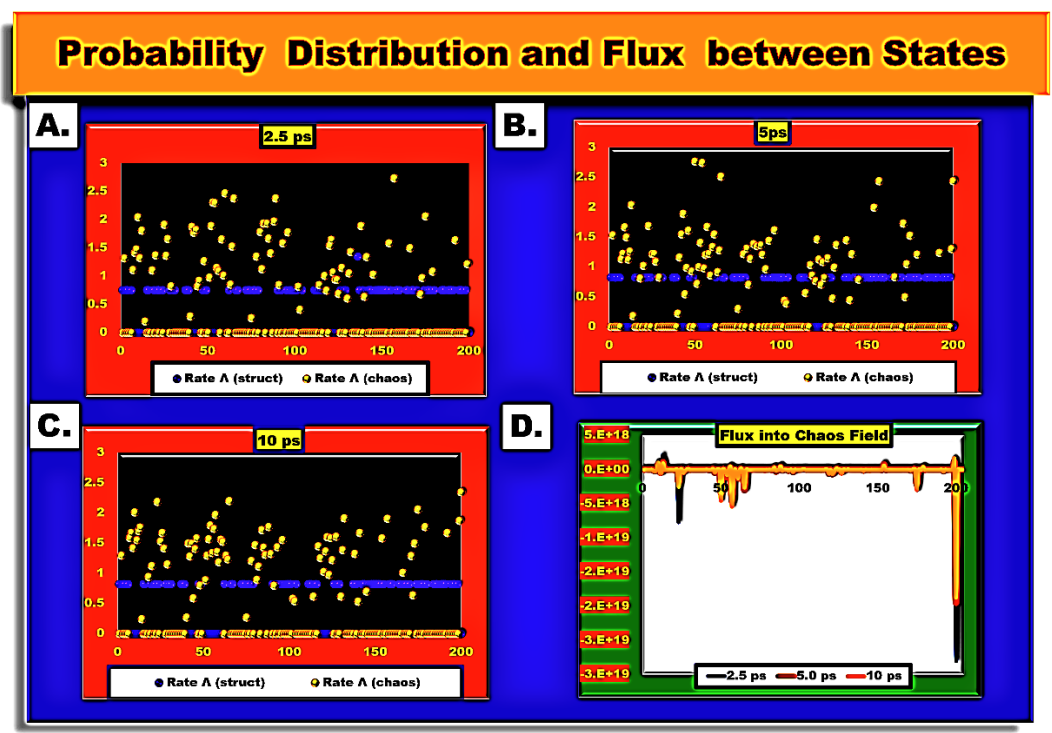

**Fig. 12** Probability Distribution and Flux at Discrete Vibration States

$$\chi(\Delta\Delta G) = \frac{\partial \chi}{\partial \Delta\Delta G} \sum_{\xi_S=1}^{n} \prod_{a=1}^{m} y_a = \int_{y=lower}^{y \leq saddle} \chi(\Delta\Delta G) dt \quad (28)$$

The summation of motions located on or below the diffusion saddle constitutes the entropy allocated to maintain the integrity of the docking protein during the conformational switch **Eq. 28**.

Eventualities of microscopic resonance damping on molecular trap stability were modeled by transmuting entropic $S = 0.5 K_B T Log(P(\Delta\Delta G))$ information to the chaos field via pressure $P(\Delta\Delta G)$ waves emanating from resonating structures. Repercussions on the chaos vector space were enumerated by predicting modifications in their motion Hamiltonians according to $H_\varphi = \Delta E - \Delta E_\Gamma$ [5]. Energy depreciation $\Delta E_\Gamma = \varphi e^{-\Gamma Z}(1 - e^{-\Gamma \tau_0})$ and evaporation of submicroscopic scale entropic distortions constituting the bulk of the molecular entanglement were captivated by applying the Boltzmann operator. Nascent resonance $\varphi$ was forecast using Le's plucked molecular string Fourier [5, 28], with oscillation period $\tau_0 = 1/\omega$ given by the inverse solution.

We discovered that the deterioration of chaotic residue motions located above the diffusion saddle occurred on a timescale of $0 - 10\ ps$. The comparatively high stability of structural residues suggests that nature designed this strategy to ensure the transient interaction and release of ligands. Multidimensional entropy flux Fig. 12d was not significantly affected by damping but only displayed modest changes across states. However, the reduction in white noise resulted in the redistribution of its probability distribution Fig. 12a-c.

*Macroscopic scalar field Damping and Trap Stability*

The docking protein is a dissipative dynamic system, thereby to model influences of macroscopic damping on molecular trap engenderment, we derived the evolution operator $\Lambda$ based on the mutation of the vibration mode $q \to (\vec{r})$ and the angular velocity $p \to \omega(\vec{r})$ Eq.29. These terms are synonymous with those given in the Navier-Stokes equation [35] and described in [36]. Whereby, we considered convolution of the continuous elastic media as the allosteric wave traverses its matrix, which entails rotation, tangential, and stretching modes of backbone residues.

$$\begin{pmatrix}\dot{q}\\\dot{p}\end{pmatrix} \to \begin{pmatrix}\dot{\vec{r}}\\\dot{\omega}\end{pmatrix} = \begin{pmatrix}\omega\\-k^0\omega\end{pmatrix} + \begin{pmatrix}0\\-\sin\vec{r}\end{pmatrix} + \begin{pmatrix}0\\\nabla H_{Ma}\cos(Z + \emptyset_0)\end{pmatrix} \quad (29)$$

$$\Lambda = Z\nabla_{\vec{r}}^2 - \overbrace{\left\{k^0\vec{r} + \sin\frac{\partial Z}{\partial \vec{r}} - \nabla H_{Ma}\cos(\omega t - \emptyset)\right\}}^{X=\{\vec{r},\omega,\emptyset\}}\frac{\partial}{\partial \omega} + \omega_0\frac{\partial}{\partial \emptyset} \quad (30)$$

Evolution operator $\Lambda$ Eq. 30 is an implicit form of the macroscopic motion equation, where state $X = \{\vec{r}, \omega, \emptyset\}$ of phase space epitomizes motions about the vibration mode, allosteric wave velocity, and energy gradient phase shifts resulting from resonance damping. Theoretically, in a continuous collision-free molecular matrix, the viscosity $-\gamma^0 \to -k^0$ term maps to spring constant $k_l(motion) = [\pi^4 n^4/l^3]_{\vec{r}^2}$ which captivates its elastic response. [5] The second term of state $X$ inspires chaos captured by angular velocity $\partial Z/\partial \vec{r}$ of the tumultuous behavior fomented by the wave. [36] The last term $\nabla H_{Ma}\cos(\omega t - \emptyset)$ portrays driving force innovations consequential to phase shift $\emptyset$ resulting from damping. Nature elegantly designed the molecular entanglement so that each of its component systems, the entropy trap, oscillatory potential, and the oscillating electromagnetic field, overlaps and interacts at state $X$. We subsequently mapped $X$ to the Hamiltonian $\Psi(t)$ by integrating and accounted for damping rate $\Gamma = \chi_{t+1}/\chi_t$ by predicting time evolution $X'(t) = Xe^{t\Lambda}$ according to Eq.31.

$$\Psi'(t) = \left\{\int k^0\vec{r} + \overbrace{\Gamma e^{t\Lambda}\sin\frac{\partial Z}{\partial \vec{r}}}^{X'(t)} - \nabla\Psi\cos(\omega t + \emptyset)\,d\vec{r}\right\} \quad (31)$$

$$= \frac{1}{2}k^0\vec{r}^2 + \Gamma e^{t\Lambda}\left(\vec{r}\sin\omega(\vec{r}) + Z\frac{\partial Z}{\partial \vec{r}}\right) - \Psi\ln[\vec{r}]\cos(\omega t + \emptyset)$$

We modeled eventualities of macroscopic field damping by allocating half the wave Hamiltonian to the surrounding fluid, thereby characterizing energy loss at the end of the conformational switch with the remainder reallocated to internal energy $\Psi$. Damped state $\Psi'(t) = \nabla\Psi - \Psi e^{-\Gamma Z}(1 - e^{-\Gamma \tau_0})$ was enumerated as before, by monitoring force modifications depicting transmutation of traversing pressure waves over the submicroscopic dimensional space. Thereby, acquiescing the discovery of damping coefficient $\Gamma$ aftermaths on state $X = \{\vec{r}, \omega, \emptyset\}$ by equilibrating the solution with Eq. 31. We delineated allosteric wave contributions to chaos vector space by predicting curl $\nabla \times \Psi'(t)$ into the thermal field followed by Boltzmann transformation to captivate the entropic distortion. Whereby, we mapped state $X = \{\vec{r}, \omega, \emptyset\} \to \chi(\Psi', t, \vec{r})$ to entropy dimensional space and actualized backbone vibration modes to chaotic state $\chi(\Psi', t, r) = \partial Y(S(\Psi', t))/\partial \vec{r}$ eventuated by standing wave dispersion $Y$ with both the nascent and innovative state transpiring consequential to operator $\Lambda = \sum_{\xi_s=1}^{n}\prod_{a=1}^{m} y_a$ Eq.32.

$$Chaos = \quad (32)$$

$$\frac{\partial}{\partial \vec{r}}\left\{\overbrace{\left(\overbrace{\frac{\partial Y(S(\Psi', t))}{\partial \vec{r}}}^{experimental}\right) - \left(\overbrace{\frac{\partial Y(S(\Psi', t))}{\partial \vec{r}}}^{predicted}\right)}^{Residual\ Y}\right\}\sum_{\xi_s=1}^{n}\prod_{a=1}^{m} y_a \Bigg|_{y>ax+bx^2+cx^3......}^{y=upper}$$

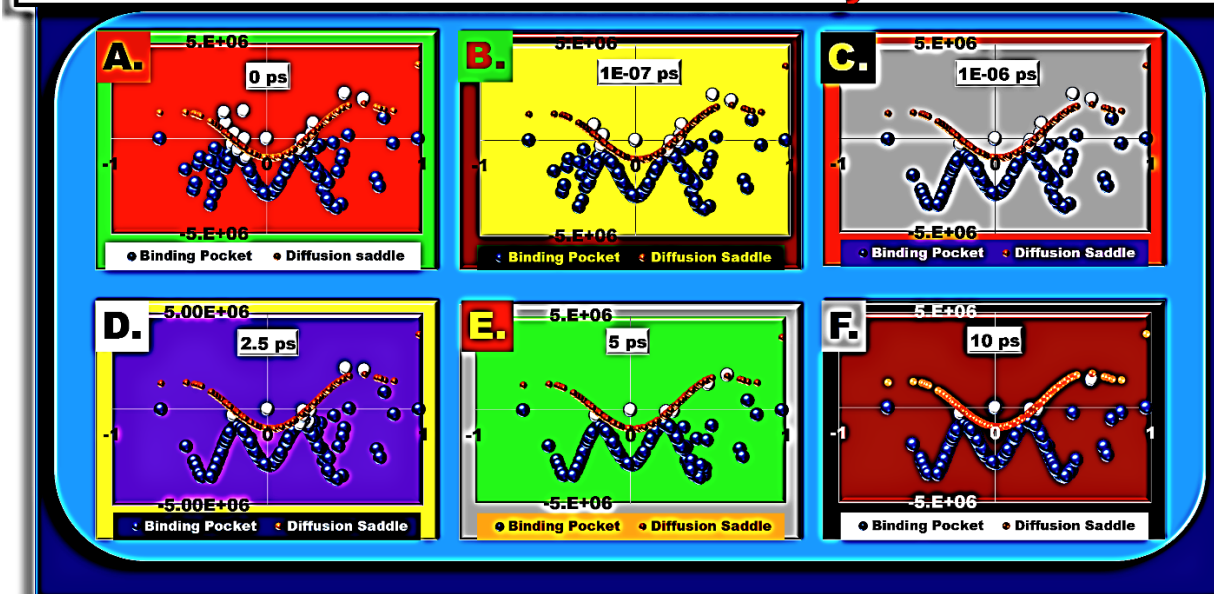

**Fig. 13** Effects of Damping on Entropic Trap Formation and Stability

Macroscopic chaos contributions to the molecular trap transpired at a significantly smaller time scale than at the microscopic scale as changes in white noise were noticeable around $t = (1e^{-07}, 2.5e^{-03})\, ps$ Fig.13a-c. Inaugural criteria for engenderment of the entropic trap and steric guidance of ligands manifested immediately following the conformational switch. At these small timescales, the standing wave fomented chaos that contributed to the lower wall and funnel of the trap. In this initial phase, white noise incited by allosteric wave turbulence underwent slow deterioration until the onset of the second phase, wherein microscopic resonance influences actualized at $t = (2.5, 10)$ picoseconds Fig.13d-f. White noise contributions from the wave significantly decreased and became almost nonexistent simultaneous to the initiation of microscopic damping and ER eventualities, whereby entropy contributions from the macroscopic field became predominantly structural. Notably, our findings imply that nature utilizes a permissibility barrier employed by the continuous elastic media's response to standing waves to synchronize macroscopic and submicroscopic communications acquiescing genomic optimization of transient ligand interaction and release.

*Effects of the EMF on the Molecular Trap*

To corroborate steric guidance by combinatorial actions of chaos, harmonic, and electromagnetic field gradients, we actualized the microscopic elastic response consequential to the dispersion factor and enacted a permissibility barrier Eq.21. We delineated allosteric wave interaction with the thermodynamic fields utilizing a 6th order polynomial regression Fig.14a. Charged residues were strongly associated with the entropic basement suggesting that the inauguration of electronic and entropic potentials coincide. Their affiliation with the aperture implies that the EMF is subject to damping but not as prone as the chaos trap. We also discovered an inverted entropy entanglement superimposing the EMF depicted by black squares that might be responsible for stabilizing the field in addition to ligand entanglement. Coexistence of these entanglements suggests an intricate level of complexity required to contrive conditions for molecular guidance and EMF stabilization.

We predicted EMF intensity $E: \{q, \omega, \vec{r}\}$ as a function of charge $q$, angular velocity $\omega$, and the vibration mode $\vec{r}$, where field fluctuations circumscribe minimum energy conformations of $\vec{r}$ microscopic states characterized by their dimensionless second moment. By projecting static electric field strengths to the docking protein topology, we discovered that the adjuvant behaviors of a weak background EMF on the dorsal side Fig.14c and strong fields affiliated with regions juxtaposing the amphipathic groove Fig.14b created an EMF gradient. The background EMF was a superimposition of the chaos field previously illustrated in Fig.5b, implying that it not only establishes the entropic potential but sets the electronic potential.

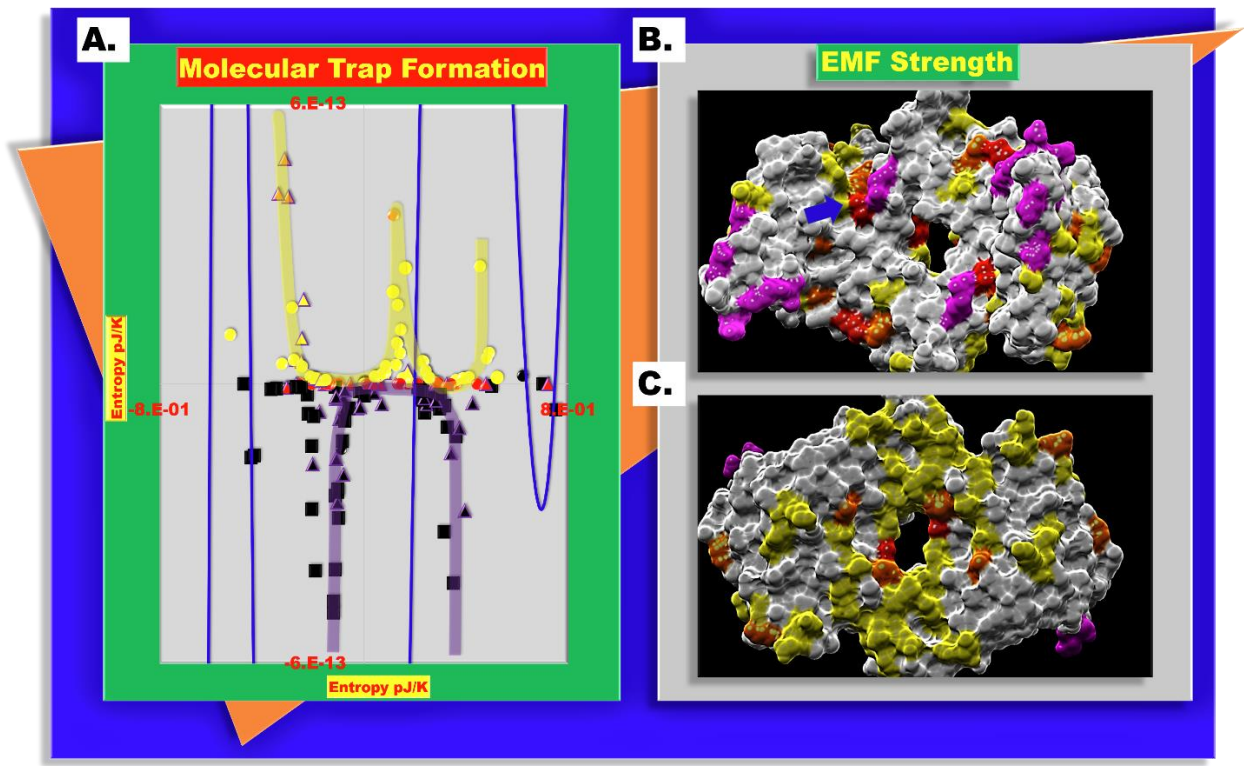

**Fig. 14 Overlap of EMF and Chaos Trap.** We analyzed the electromagnetic field (EMF) fomented by resonances of charged residues as a function of their state $E:\{q,\omega,\vec{r}\}$. Overlap of entropic and EMF entanglements, the standing wave is depicted by the blue curve, charged residues are identified as triangles, structural residues squares, and non-charged chaotic motions by circles (**A**). Projection of EMF to docking protein structure, intensity increases $E:\{yellow \rightarrow orange \rightarrow red \rightarrow violet\}$ (**B, C**). Arrow depicts overlap of EMF with the fusicoccin and molecule (2S) - (2) - methoxyethyl pyrrolidine binding site.

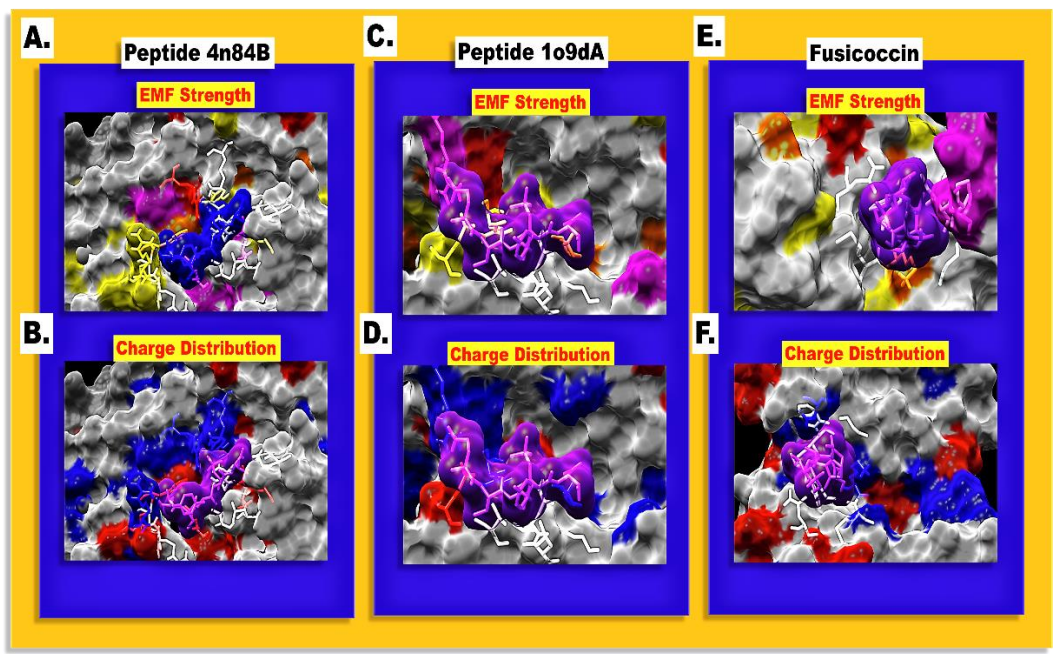

**Fig.15 The Role of the EMF in steric Guidance and Ligand Binding.** We analyzed the electromagnetic field fomented by resonances of charged residues as a function of their state $E:\{q,\omega,\vec{r}\}$, intensity increases $E:\{yellow \rightarrow orange \rightarrow red \rightarrow violet\}$. To demonstrate the role of the EMF, we characterized the binding of peptide 4n84B **Fig. A** (blue)**,** peptide 1o9dA **Fig. C** (magenta), and fusicoccin **Fig. E** (magenta). We also illustrated the charge distribution about these molecules in their perspective ligand binding sites, $blue = (+)$, $red = (-)$, **Figs. B, D, F**. Peptide 4n84B it was shown in magenta in **Fig. B**.

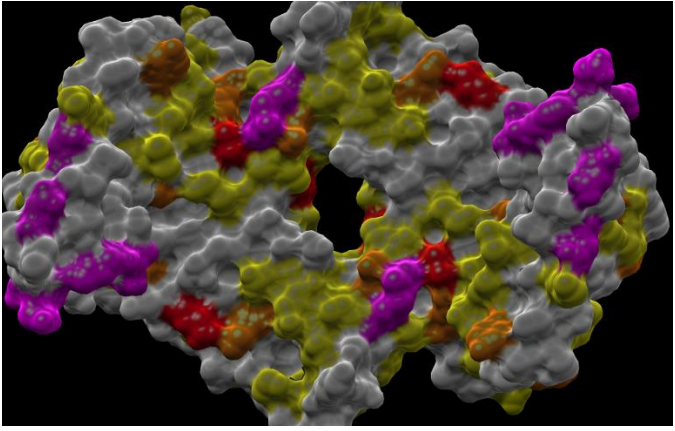

**Fig.16 Electromagnetic Field Gradient**

Strong EMFs were also associated with binding sites within the groove.

To gauge the conductivity of the amphiphilic groove interior space to EMF propagation, we predicted water exclusion by analyzing amino acid hydrophobicity utilizing ProtScale [66]. We discovered solvent exclusion wells localized within the groove and simulated vacuum conditions that encompassed a magnetic permeability approaching an open space maximum limit $\mu_0 \Rightarrow 4\pi \times 10^{-7}\ Hm^{-1}$. Thereby, the space must also comprise a prominent electric permittivity approaching the limit $\varepsilon_0 = 1/\mu_0 c^2$ that assuages the background EMF forward propitiating reinforcement of local fields affiliated with ligand-binding sites Fig.16.

$$\mathcal{E} = \left\{ \int_{\partial V} dS \int_{\rho(0)}^{\rho(1)} \nabla\varphi \cdot d\rho(\lambda) \right. \\ \left. + \int_{\partial S} dl \int_{\sigma(0)}^{\sigma(1)} \nabla\varphi \cdot d\sigma(\lambda) \right\}_{X(\Gamma)} \quad (33)$$

EMFs magnetically align molecules by actualizing a traverse force $\vec{F} = q(\vec{E} + \vec{v} \times \vec{B})$ onto velocity vectors dependent on electric field strength $\vec{E}$, velocity $\vec{v}$ of the molecule acquiring the molecular trap, and magnetic flux $\vec{B}$. The ligand also displays a charge distribution whose static electric field counteracts the docking protein, whereby its resonance foments an oppositional gradient. Aperture and background EMFs genesis an electromotive force gradient $\mathcal{E} = \vec{\mathcal{E}}_{aper} - \vec{\mathcal{E}}_{bg}$ responsible for ligand entanglement within a multidimensional potential that introduces and pulls them down an energetic funnel constituted of superimposed, chaotic, harmonic, and electromagnetic fields into the groove. Where localized EMFs reorientate, sequester, and lock them into place Fig. 15. We noticed that EMFs localized specifically in binding sites and that charge distribution was site-specific.

Our model predicts electromotive force coincidental to EMF components of the molecular trap consonant to Krishtalik. [58, 59] Electric field $\nabla\varphi = \vec{E}$ actualizes consequential to a potential gradient $\nabla\varphi$ created by counteracting ligand and docking protein EMFs. During the acquisition of the molecular trap, an electromotive force ensues from the reinvented field characterized by its curl $\nabla \times \vec{E} = f\left\{\oint_{\partial V} \vec{E} \cdot dS, \oint_{\partial S} \vec{E} \cdot dl\right\}$ over the enclosed area of the protein Eq.33. The second term captivates localized currents fomented by communicating residues that emit magnetic fields responsible for orientating ligands in their respective binding sites whose magnetic flux $\vec{B} \equiv \vec{E}$ is equivalent to the electric fields engendered. EMFs dynamically regenerate and metamorphose as volume $V$ and surface areas $S$ undergo expansion and contraction responsive to vibration mode fluctuations. EMF gradient innovations transpire contemporaneously to resonance states and modulate in conformity to damping coefficients. Electromotive field deterioration propagates with irradiation into the interior space. Whereby EMF intensity $I_{rms} = \mathcal{E}/4\pi r^2$ is consequential to prolongation of the radius into free space.

*Effects of the Harmonic gradient on the Molecular Trap*

We substantiated vibrational ramifications on molecular entanglement conception respective to chaos and electromagnetic field gradients by illustrating the harmonic entanglement $¥ = \partial\Delta\Delta G/\partial\psi$ is inaugurated by coadjuvant influences of resonance $\psi$ and allosteric free energy $\Delta\Delta G$ scalar fields. Our findings signify that structural harmonics acquiesce consequential to the available $\Delta\Delta G$ and attenuate approaching a limit $\Delta\Delta G = 0$ where free energy is exhausted Fig. 17d. Comparatively, $\Delta\Delta G$ landscapes appear more ordered than chaos fields as structural internal energy $\partial\Psi = \partial Q \rightleftarrows \partial W$ equilibria is shifted toward continuous elastic media organization versus disorganization Fig.17c. Consequently, disruption translations into the submicroscopic dominion insinuate chaos distribution is not entirely incidental. As attributes eventuated by the First Law of Thermodynamics in the higher dimensions materialize as hotspots scattered throughout the disruption field. Organizational retention across Hamiltonians implicates the $\Delta\Delta G$ vector space as a necessary influencer in the organization of tumultuous motions into an oscillatory trapping apparatus. Whereby, harmonic trap genesis transpires by congruent activities of the disruption fields. The chaos vector space wrapped concentrically around the vibration mode depicting circumvention of turbulence bout the wave. The $\Delta\Delta G$ configuration was orthogonal Fig.17d and a consequence of their resultant Fig.17b. Compulsorily, the conjoined actualizations of these thermodynamic drivers created an oscillatory entanglement in the direction of the allosteric wave orthogonal to the vibration mode Fig.17d.

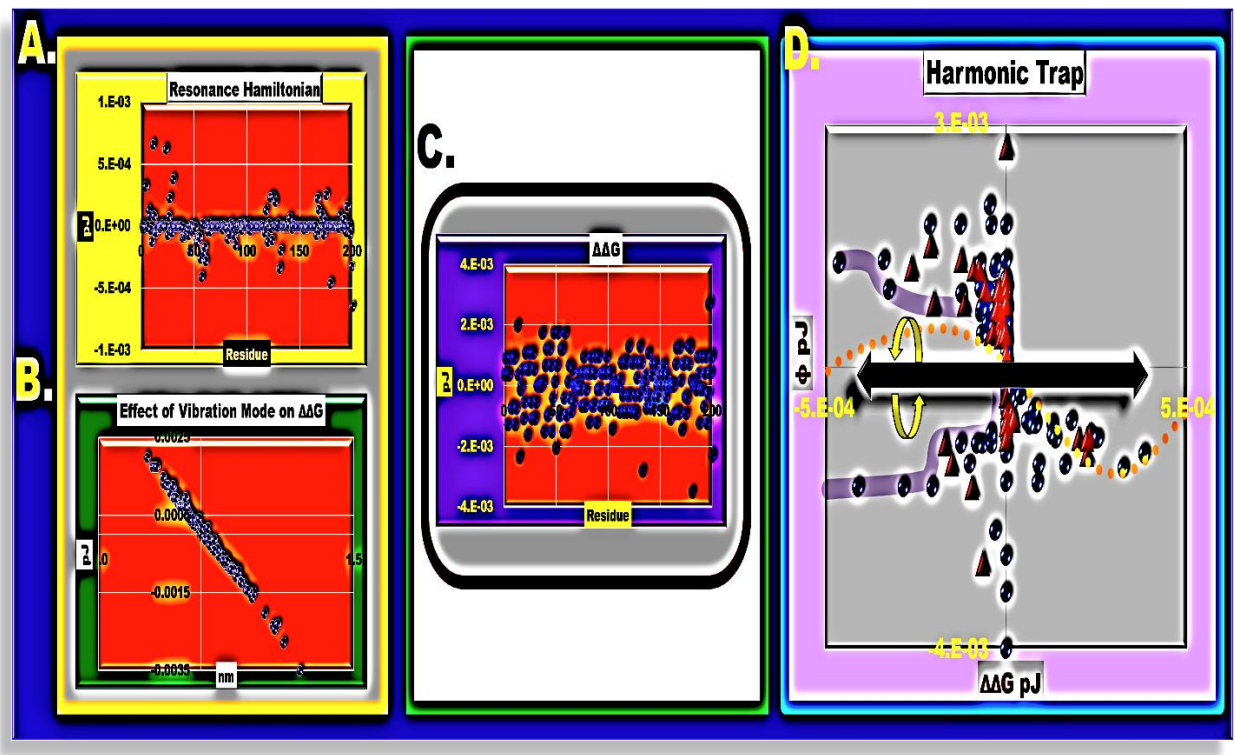

**Fig.17 Formation of the Harmonic Trap.** Charged residues are depicted as red triangles and allosteric standing wave orange dots.

$$£ = \left\{ \int_{\partial V} dS \int_{\rho(0)}^{\rho(1)} \nabla \mathcal{H} \cdot d\rho(\Psi) + \int_{\partial S} dl \int_{\sigma(0)}^{\sigma(1)} \nabla \mathcal{H} \cdot d\sigma(\Psi) \right\}_{X(\Gamma)} \quad (34)$$

Intimate overlaps of harmonic and electromagnetic entanglements imply preeminent realizations of the EMF transpire at the microscopic scale Fig.17d. Whereby oscillatory field intensity is an actuality of residue spatial dispersion and resonance respective to the available allosteric free energy. Consequently, harmonic state $£$ was modeled synonymously to chaotic state $§$ by considering harmonics over closed volume and surface area of the docking protein Eq. 34. Driving force $F(\chi) \to \nabla \mathcal{H} = \nabla \psi + \nabla(\Delta\Delta G)$ of chaotic states that spawn vibrational motions are an actualization of resonance $\psi$ and $\Delta\Delta G$ gradients which procreate a concentric field that sterically guides ligands. Once entangled, downward forces induce molecules into an oscillatory potential. The harmonic entanglement's firmness is a function of the spatial distribution of internal energy respective to the closed volume $\rho(\Psi)$ and surface $\sigma(\Psi)$. Internal energy $\Psi$ encaptivates the dynamic potential of the trapping apparatus. Notably, while harmonic basement residues exhibit less anarchic motions, they subsist at significantly higher probability densities Fig.17a and gyrate at higher frequencies, acquiescing entrapment of ligands once binding sites are procured. Local intensities $I = £/4\pi r^2$ materialize across the entanglement as a function of the instantaneous radiation radius and dynamic localized fields that modulate over time due to fluctuations in the vibration mode happenstance of damping.

## Concluding Remarks and Commentaries

We incontrovertibly substantiate the existence of an innate molecular trap that promotes ligand binding, selection, and electronic guidance into binding sites located within the 14-3-3 ζ amphipathic groove. The molecular entanglement transpires coincidental to disruption transferences across time-spatial dimensions eventuated by the tumultuous behavior of allosteric standing waves during propagation of the continuous elastic media. The trapping apparatus encompasses chaotic, harmonic, and electromagnetic entrapment systems comprising disparate physics that genesis contemporaneously and interconnect at their motion states. Significantly, our discoveries imply genome supervision of the transient interaction and release of molecules based on the organization of white noise emitted by residue vibration modes, thereby acquiescing manipulation of the Second Law of Thermodynamics by fulfilling entropy requirements while simultaneously ascertaining a significant degree of temporary structure.

Significantly, entanglement traps discovered herein share many reported features of black holes that distort space-time, acquiescing a bulk constituting an induction well or event horizon and a funnel in higher-dimensional space. [67] These structures form by collapsing stars creating an entropic entanglement. [68-71] Congruous events occur in docking proteins at the conformational switch actualizing a bulk constituting an aperture and funnel, which converge on a stable entropic basement associated with the allosteric wave synonymously to a convergence of distorted space-time at the singularity. The molecular trap materializes in the submicroscopic dimension as a tidal-like response to the standing wave that dictates entropy flow and dimensional warping. The response is consequential to the protein's internal structure and residue density, as we illustrated in Eq. 12-14, 33, 34. And as demonstrated in comparisons of the native and synthetic entanglements, bulk distortion is a function of the wave's energy density. Similar physics transpire in scalar objects actualized by deformations due to gravity waves refs. [72-74].

In 14-3-3 ζ docking proteins, a rapid contraction at the conformational switch from the 'open' to 'closed' conformational state eventuates a multidimensional trapping apparatus. The resulting contractile forces generate dimensional distortions accompanied by a rolling macroscopic entropy through the protein fabric. Residual domain wobbling and helix oscillations coalesce the distortion fields. Notably, a contemporaneous and none related study of actomyosin remodeling by Al-Izzi [75] demonstrated vortex genesis during contractile motions. Our findings suggest similar events transpire in docking proteins. However, due to synchronization to signal transduction occurs at a much smaller time scale.

Like their enormous relatives, bulk entropy increases with radius Fig. 6a,b dependent on microstates depicting innovations in the motion equation [76]. Whereby, the continuous media's elastic response to the allosteric wave mitigates the organization of entropy analogous to space-time responses to their stellar cousins. Saliently, both apparatuses are entropic devices materialized by the organization of disruption fields into temporary structures, dependent on the reference time frame. [77-81] Similarly to black holes, these miniature objects comprise multi-dimensional entropy, interacting and superimposed harmonic and electromagnetic fields that supervise coincidental interactions with passing objects or molecules induced into chaotic entanglement. [82-85]

Notably, findings reported herein imply that nature advantageously manipulates entropy from microscopic to infinitely large scales to engineer efficient information processes. It further suggests the existence of an entropy-based fundamental evolution force responsible for the universal blueprint that is repetitious down to lower thermodynamic scales. Such molecular phenomena open unique opportunities for the resolution of classical and quantum physics. While our focus was not astrological, as a side note, there is also an unavoidable implication that a higher level of structure is associated with stellar objects such as black holes.

### Acknowledgements
I would like to extend gratitude to Eva Davis and Hannah Davis for support, and to the Texas Advanced Computing Center (TACC) at the University of Texas.